\documentclass[apj]{emulateapj}

\newcommand{\twom}{2M1207b}
\newcommand{\hrb}{HR8799b}

\newcommand{\azero}{${\rm a}_{0}$}
\newcommand{\teff}{${\rm T}_{\rm eff}$}
\newcommand{\pmin}{${\rm P}_{\rm min}$}
\newcommand{\kzz}{${\rm K}_{\rm zz}$}
\newcommand{\logg}{$\log(g)$}
\newcommand{\rjup}{${\rm R}_{\rm Jup}$}
\newcommand{\mjup}{${\rm M}_{\rm Jup}$}
\newcommand{\lobs}{${\rm L}_{\rm obs}$}
\newcommand{\phoenix}{{\tt PHOENIX}}

\makeatletter

\newcommand{\Rmnum}[1]{\expandafter\@slowromancap\romannumeral #1@}
\makeatother

\shorttitle{The Atmosphere of HR8799 \MakeLowercase{b}}
\shortauthors{Barman et al.}

\begin{document}
\bibliographystyle{apj}

\title{Clouds and Chemistry in the Atmosphere of Extrasolar Planet HR8799 \MakeLowercase{b}}
 \author{Travis S. Barman}
 \affil{Lowell Observatory, 1400 W. Mars Hill Rd., Flagstaff, AZ 86001, USA\\ 
        Email: {\tt barman@lowell.edu}}

 \author{Bruce Macintosh, Quinn M. Konopacky}
 \affil{Lawrence Livermore National Laboratory, 7000 East Avenue, Livermore, CA 94550, USA}

 \author{Christian Marois}
 \affil{National Research Council Canada, Herzberg Institute of Astrophysics, 
        5071 West Saanich Road, Victoria, BC V9E 2E7, Canada}

\keywords{planetary systems - stars: atmospheres - stars: low-mass, brown dwarfs}

\begin{abstract}
Using the integral field spectrograph OSIRIS, on the Keck \Rmnum{2}\ telescope, broad
near-infrared $H$ and $K$-band spectra of the young exoplanet \hrb\ have
been obtained.  In addition, six new narrow-band photometric measurements have
been taken across the $H$ and $K$ bands.  These data are combined with
previously published photometry for an analysis of the planet's atmospheric
properties.  Thick photospheric dust cloud opacity is invoked to explain the
planet's red near-IR colors and relatively smooth near-IR spectrum.  Strong water
absorption is detected, indicating a Hydrogen-rich atmosphere.  Only weak
CH$_4$ absorption is detected at $K$ band, indicating efficient vertical mixing
and a disequilibrium CO/CH$_4$ ratio at photospheric depths. The $H$-band
spectrum has a distinct triangular shape consistent with low surface gravity.
New giant planet atmosphere models are compared to these data with best fitting
bulk parameters, \teff\ = 1100K $\pm 100$ and $\log(g) = 3.5 \pm 0.5$ (for
solar composition).  Given the observed luminosity ($\log L_{\rm obs}/L_{\odot}
\sim -5.1$), these values correspond to a radius of 0.75 \rjup $^{+0.17}_{-0.12}$
and mass $\sim$ 0.72 \mjup $^{+2.6}_{-0.6}$ -- strikingly inconsistent with
interior/evolution models.  Enhanced metallicity (up to $\sim$ 10 $\times$ that
of the Sun) along with thick clouds and non-equilibrium chemistry are likely
required to reproduce the complete ensemble of spectroscopic and photometric
data {\em and}  the low effective temperatures ($<$ 1000K) required by the
evolution models.  

\end{abstract}
\vspace{6pt}

\section{Introduction}

Direct imaging of exoplanets is now becoming a regular occurrence with the
recent discoveries of \twom\ \cite[]{Chauvin2005}, Fomalhaut b
\cite[]{Kalas2008}, Beta Pic b \cite[]{Lagrange2009}, 1RXS J1609 b
\cite[]{Lafreniere2008}, and the quadruple-planetary system HR8799 b, c, d,
and e \cite[]{Marois2008,Marois2010}.   HR8799 is, at present, a rare find
among exoplanets with its four potentially massive planets ($\sim$ 5 -- 10
M$_{\rm Jup}$) orbiting a $\sim~30$ Myr A5 star \cite[]{Zuckerman2011}, and is
the first and, so far, only multi-planet system to be directly imaged.  All
four planets are accessible to photometric and spectroscopic followup
observations across many wavelengths and, in this paper, near-IR spectra of
\hrb\ are folded into a large set of photometric data for a broad-band
analysis of the planet's photospheric properties.

Since its discovery, a number of new photometric and astrometric measurements
of the HR8799 system have been made.  These new data have broadened the
photometric wavelength coverage and extended the time coverage of orbital
motion \cite[]{Metchev2009, Lafreniere2009, Hinz2010, Currie2011}.  The very
precise astrometric data have already inspired a number of papers focused on
the orbital dynamics and formation of the system; however, the conclusions
about the planetary masses and long term stability of the system from these
works are mixed \cite[]{Fabrycky2010, Reidemeister2009, Gozdziewski2009,
Dodson-Robinson2009}.  As this paper focuses on photospheric properties, the
dynamical nature of the system will not be discussed here.  

The goal of this paper is to characterize the properties of \hrb,
independent of what interior/evolution models predict and independent of what
the age or dynamics of the system suggest.  Following a description of the new
spectra and photometry added by this work (Section 2), the available set of
photometric and spectroscopic data are compared to similar data for brown
dwarfs -- the closest analogues to young giant planets (Section 3).  The
atmospheric cloud and chemical properties of \hrb\ are inferred by comparing
a new set of substellar atmosphere models to the available set of photometry and
spectroscopy (Sections 4 and 5). The results are compared to the predictions of
interior/evolution models.  Also, given the potential for similarities, a
comparison is made between \hrb\ and \twom\ (Section 6).  The conclusions
from this work are summarized in the last section.

\section{Observations and Data Reduction}

\subsection{OSIRIS Spectroscopy}

\begin{figure}[t]
\plotone{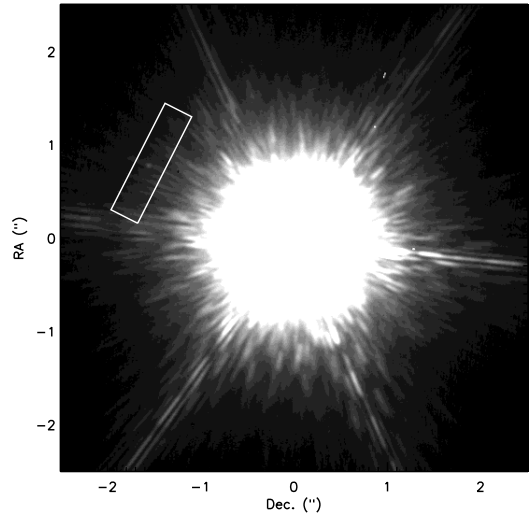}
\caption{An unprocessed NIRC2 image (20 second exposure) of HR8799.  The
rectangle shows the orientation and size of the OSIRIS spectrograph FOV
(0.02\arcsec\ scale). \hrb\ is barely visible in the center of this rectangle
and comparable in brightness to the speckle contamination.  Dithering was done
along the long-axis of the FOV.
\label{fig1}}
\end{figure}

Observations of \hrb\ were made on July 22, 23, and 30 (UT) of 2009 and on
July 11 and 13 (UT) of 2010 at the W. M. Keck Observatory. Natural guide star
adaptive optics and the OH-Suppressing Infra-Red Imaging Spectrograph (OSIRIS)
instrument \cite[]{Larkin2006} were used on the Keck \Rmnum{2}\ telescope.  Broad-band
$H$ and $K$ spectra were obtained over a 0.32 $\times$ 1.28\arcsec\ patch of
the sky surrounding the planet at 0.02\arcsec~spaxel$^{-1}$ (see Fig.
\ref{fig1}), where spaxel refers to one OSIRIS spatial resolution element.
OSIRIS is an integral field spectrograph that uses a grid of lenslets in the
focal plane to dissect the field of view (FOV). The light from each lenslet is
dispersed by a grating, producing an array of spectra that are reassembled in
the data reduction pipeline (DRP) to a [x, y, $\lambda$] data cube.  A series
of 21 $H$-band and 18 $K$-band 900 second exposures were obtained.  Appropriate
dark, sky, and A0 telluric standard star observations were also obtained. See
Table \ref{tab1} for a summary of these observations.

\begin{deluxetable}{lcccccccc} 
\tablecolumns{8} 
 \tablewidth{0pt} 
\tablecaption{OSIRIS Observations} 
\tablehead{\colhead{target}& \colhead{\#exp} & \colhead{exp. time (s) }&\colhead{band pass} & \colhead{UT date}}
\startdata 
\hrb          & 9          & 900        & $K$      & 07-22-09 \\
HD210501         & 5          & 30         &          &          \\
BD+14 4774       & 4          & 30         &          &          \\
                 &            &            &          &          \\ 
\hrb          & 6          & 900        & $H$      & 07-23-09 \\
HD 210501        & 7          & 30         &          &          \\
BD+14 4774       & 5          & 30         &          &          \\
                 &            &            &          &          \\ 
\hrb          & 7          & 900        & $H$      & 07-30-09 \\
BD+14 4774       & 7          & 30         &          &          \\
                 &            &            &          &          \\
                 &            &            &          &          \\
\hrb          & 9          & 900        & $K$      & 07-11-10 \\
HD210501         & 3          & 30         &          &          \\
                 &            &            &          &          \\
\hrb          & 8          & 900        & $H$      & 07-13-10 \\
HD210501         & 3          & 30         &          &          \\
\enddata 
\label{tab1}
\end{deluxetable}

The OSIRIS DRP was used to produce basic calibrated data (BCD), and includes
spatial rectification of the raw data, wavelength and dispersion solutions, sky
subtraction, cosmic ray rejection, and so forth \cite[]{Krabbe2004}.  On all
three nights in 2009, OSIRIS was operating above its normal detector
temperature ($\sim 68$K). On July 22, the temperature was $\sim$ 79K, but
holding steady.  On July 23, the temperature was $\sim$ 80K and also stable
throughout the night.  However, on July 30, the temperature was not stable and
increased from 83.56K at the beginning of the observations to 84.54K at the
end. During this period of unstable temperatures, the detector noise in the
dark frames increased exponentially.  To avoid taking 900 sec sky and dark
frames repeatedly throughout the night, ``scaled darks" were generated for each
900 sec exposure on the science target by scaling actual darks taken at the
beginning and end of the half-night. An exponential function was fit to the
total counts in each actual dark from all three nights and, using this
function, scaled darks were generated for the science-target cubes on July 30.
In 2010, OSIRIS was functioning perfectly, and the observing conditions were
excellent.

The final BCD data cubes were, at every spatial position, divided by the
spectrum of an A0 standard star and multiplied by a black body ($T = 9600$K).
The spectrum of the A0 star was extracted from its data cube by fitting a
2D-Gaussian at every monochromatic image-slice, rather than using the
fixed-aperture extraction routine in the OSIRIS DRP. BD+14 4774 was discovered
to be a binary (0.07\arcsec\ separation) with a likely mid to late-type companion;
for this star a two-component Gaussian PSF was used to extract the A0 fluxes.
The final BCD are cubes with wavelength and two spatial dimensions.  At both $H$
and $K$ bands, the planet was clearly visible in all median-collapsed cubes
(median along wavelength).  The nominal resolving power of OSIRIS is $R \sim
4000$ with 1600 wavelength channels.  However, to improve the signal-to-noise
ratio, the BCD cubes were binned to 25 wavelength channels, down to $R \sim
60$.  The wavelength positions of OH sky-lines and telluric star features were
measured and no artificial wavelength shifts were discovered; a potential
concern given the temperature variations encountered during the observations
\footnote{Several updates of the OSIRIS DRP became available during the course of
this work; one of these was directly related to preventing unwanted wavelength
shifts ({\tt http://irlab.astro.ucla.edu/osiris/}).}.

\begin{figure}[t]
\plotone{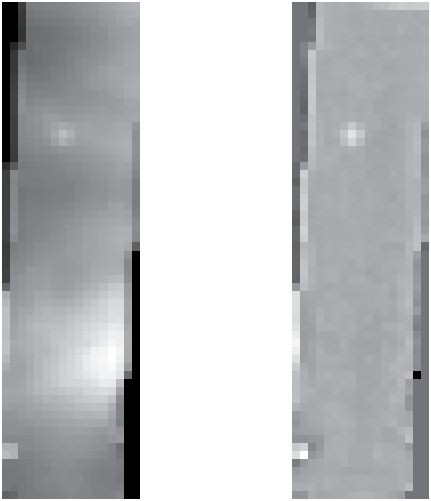}
\caption{{\em Left}: A single $H$-band basic calibrated data cube (900s
exposure at one dither position), median combined along the wavelength
dimension.   The planet is visible in the upper left quadrant. Bright
speckles are visible across most of the image.  The image is oriented
such that the star is to the right. {\em Right}: Same data as in the
left panel, but after speckle-suppression.
\label{fig2}}
\end{figure}

While the planet is easily detected in all of the BCD cubes, contamination by
scattered star light remains a concern.  Figure \ref{fig1} shows the FOV and
orientation for the OSIRIS observations and the extent of speckle contamination
in the observed region.  Since the spatial scaling of the speckles, radially
from the star, is proportional to $\lambda$, speckles can intersect the spatial
location of the planet in a wavelength-dependent manner.  To suppress the
speckle pattern, each BCD cube was processed with custom IDL routines that
first subtracts a (spatially) low-pass filtered version of the image slice in
each wavelength channel and rebins the data cube (by taking the median in the
$\lambda$ dimension) to 25 wavelength channels (R $\sim$ 60).  Each
monochromatic image-slice of this rebinned cube is magnified about the stars
position by $\lambda_{m}/\lambda$, where  $\lambda_{m}$ is the median
wavelength in either $H$ or $K$ band.  In these shifted data cubes, individual
speckles align and are fit with simple polynomial functions of $\lambda$.
Prior to this step, the planet is masked to avoid biasing the polynomial fit.
The final polynomial fits to the speckle fluxes are subtracted from the shifted
cube before finally reversing the magnification of the cube, placing the planet
back in its original position.  Collapsed $H$-band data cubes, before and after
speckle removal, are shown in Fig. \ref{fig2} (a similar improvement is seen
in the before and after $K$-band data cubes).  The speckle suppression scheme
described here is similar to that used for the low-mass companion GQ Lupi b
\cite[]{McElwain2007}; however, McElwain et al.  did not rescale the cubes for
speckle fitting but instead fit the underlying speckle halo in each spectral
bin of the data cube since each spatial location is more than critically
sampled.

\begin{figure*}[!thb]
\plotone{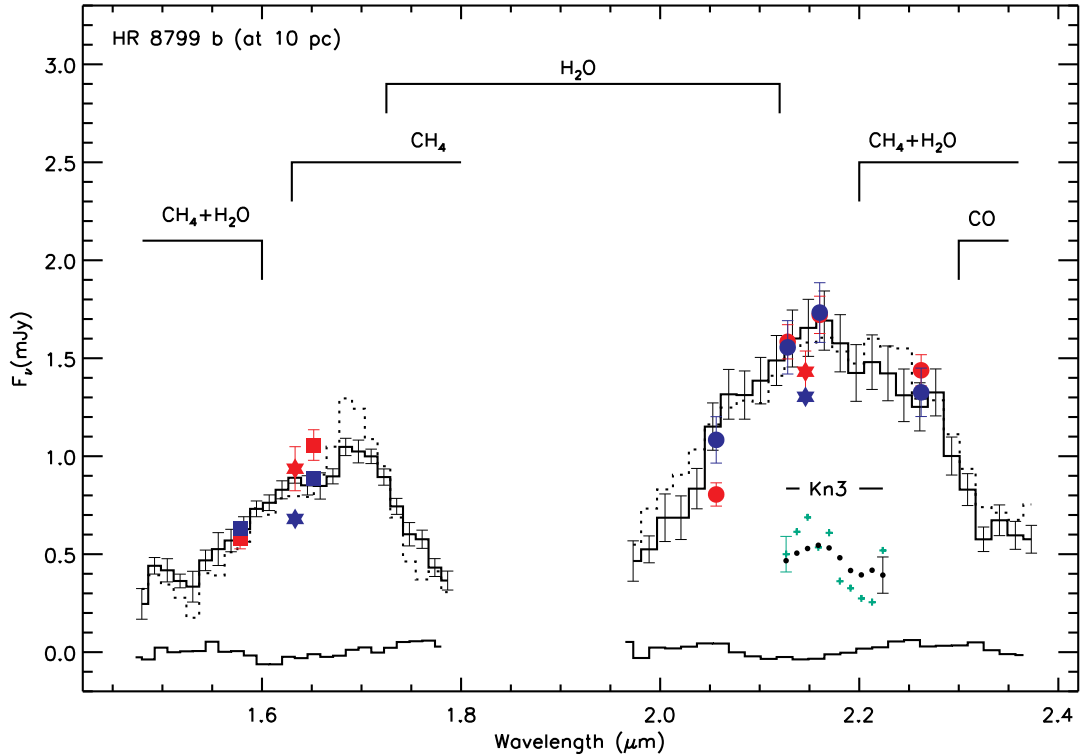}
\caption{OSIRIS $H$ and $K$-band fluxes of \hrb\ (scaled to 10 parsecs)
plotted with 1-$\sigma$ uncertainties.  The location of prominent water,
CH$_4$, and CO absorption bands are indicated. The fluxes extracted without
the speckle suppression algorithm are shown as dotted lines.  The bottom two
curves are the mean residuals of fake planets with flat spectra extracted from
the same data cubes (see text for details).  The Kn3 spectrum of
\cite{Bowler2010} is shown as green pluses (scaled arbitrarily down) over
plotted with the broad-band spectrum (black points) interpolated onto the
Bowler et al. Kn3 wavelength points. The mean 1-$\sigma$ uncertainties across
the Kn3 range are shown at either end for each data set.  The larger red filled
symbols are the NICI CH$_4$ short/long (boxes), NIRC2 narrow (circles), and
NIRC2 broad-band (stars) photometry taken from Table \ref{tab3}.  Blue symbols
are the corresponding photometry derived from the OSIRIS spectra.
\label{fig3}}
\end{figure*}

The speckle suppression steps described above contain two important variables:
the size of the masked region (centered on the planet) and the order of the
polynomial used to fit the speckle fluxes.  To determine the best pair of
parameters, ten fake planets (similar in brightness and FWHM as the real planet
but with constant F$_\lambda$) were inserted into the BCD data cubes, evenly
spaced but avoiding the real planet location as well as the edges of the FOV.
These cubes were processed using the twelve combinations of four mask radius
values (from 0.5 to 2 in 0.5 spaxel steps) and three polynomial orders (1, 2
and 3).  The parameters resulting in extracted fake planet spectra with the
smallest mean deviation from a flat line and smallest RMS($\lambda$) were mask
radii equal to 1.5 and 2.0 for $H$ and $K$ bands respectively and a polynomial
order equal to 1 for both bands.   The mean residuals of all ten fake planet
spectra are shown in Fig.  \ref{fig3}, straddling zero flux.  The
RMS($\lambda$) for the fake planet spectra were comparable to, or smaller than,
the final uncertainties adopted for the planet spectrum, providing confidence
that the errors have not been significantly underestimated.  Furthermore, the
data collected in 2009 compare very well to those from 2010, indicating that
the thermal instabilities and scaled darks had little impact on the final
reduced spectra.

The planet fluxes were obtained from the speckle-suppressed data (SSD) cube by
convolving each monochromatic image-slice with a two-dimensional, Gaussian,
point-spread-function with FWHM = 2.2 ($H$-band) and 2.7 ($K$-band) spaxels.
The final spectra were flux calibrated using $H$-band \cite[]{Metchev2009}
and a revised $K_s$ magnitudes (discussed below).  The final extracted spectra
and corresponding error-bars (taken as the RMS of all the extracted spectra in
each band-pass) are shown in Fig. \ref{fig3} and listed in Table
\ref{tab2}.

As can be seen from the residuals of the fake planet fluxes (Fig.
\ref{fig3}), the noise in the spectra is correlated between different
wavelengths.  This is a natural consequence of the fact that the noise is
primarily from the speckle pattern - the typical speckle has a monochromatic
size of $\lambda$/D, as does the extraction box for the planetary spectrum. A
single speckle's position scales with wavelength proportional to $\lambda$. At
$K$ band, the planet is located approximately 42 $\lambda$/D from the star. For
the speckle to move completely across the extraction box requires a change in
wavelength of (43/42-1) $\times$ 2.1 $\mu$m = 0.05 $\mu$m.  Hence speckle noise
patterns will be correlated across the spectrum over a range of 0.05 $\mu$m
\cite[]{Sparks2002}. In principal this should be formally accounted for in the
estimate of $\chi^2$ when fitting models to the data (described below).
However, since the spectral features being fit are also of width $\sim$0.05
$\mu$m or larger, the correlations in the noise have been ignored.

The planet spectra, before and after speckle suppression, are also compared in
Figure \ref{fig3}.  Near 1.7 $\mu$m, the impact of a bright speckle can
be seen as a fairly sharp peak spanning two wavelength channels.  Speckle noise
and photon noise are roughly equal in these data sets at this binning, so
speckle suppression improved the SNR by approximately 50\%. Most importantly,
the speckle noise that was removed was spectrally correlated across multiple
channels and non-Gaussian, e.g. the feature at 1.7 $\mu$m in Figure
\ref{fig3}, and would have a much more significant effect on the final
interpretation than the white photon noise.

\begin{deluxetable}{lcccccccc}
\tablecolumns{7}
 \tablewidth{0pt} 
\tablecaption{OSIRIS $H$ and $K$ Spectra (scaled to 10 parsecs)}
\tablehead{\colhead{$\lambda$}& \colhead{$F_\nu$ (mJy)} & \colhead{error (1-$\sigma$)} &
\colhead{ } &   \colhead{$\lambda$}& \colhead{$F_\nu$ (mJy)} & \colhead{error (1-$\sigma$)}}
\startdata
  1.48 & 0.25 & 0.08 &  & 1.97 & 0.47 & 0.10 \\
  1.49 & 0.44 & 0.04 &  & 1.99 & 0.52 & 0.07 \\
  1.50 & 0.42 & 0.06 &  & 2.00 & 0.69 & 0.12 \\
  1.52 & 0.36 & 0.03 &  & 2.02 & 0.69 & 0.11 \\
  1.53 & 0.33 & 0.08 &  & 2.04 & 0.83 & 0.10 \\
  1.54 & 0.47 & 0.05 &  & 2.05 & 1.15 & 0.12 \\
  1.56 & 0.53 & 0.08 &  & 2.07 & 1.32 & 0.13 \\
  1.57 & 0.57 & 0.06 &  & 2.08 & 1.31 & 0.12 \\
  1.58 & 0.63 & 0.06 &  & 2.10 & 1.39 & 0.12 \\
  1.59 & 0.73 & 0.04 &  & 2.12 & 1.49 & 0.13 \\
  1.61 & 0.76 & 0.04 &  & 2.13 & 1.60 & 0.15 \\
  1.62 & 0.83 & 0.05 &  & 2.15 & 1.66 & 0.15 \\
  1.63 & 0.89 & 0.03 &  & 2.16 & 1.69 & 0.15 \\
  1.65 & 0.85 & 0.05 &  & 2.18 & 1.58 & 0.15 \\
  1.66 & 0.85 & 0.07 &  & 2.20 & 1.42 & 0.14 \\
  1.67 & 0.90 & 0.04 &  & 2.21 & 1.48 & 0.14 \\
  1.68 & 1.05 & 0.04 &  & 2.23 & 1.42 & 0.14 \\
  1.70 & 1.02 & 0.06 &  & 2.24 & 1.31 & 0.13 \\
  1.71 & 1.00 & 0.04 &  & 2.26 & 1.25 & 0.12 \\
  1.72 & 0.89 & 0.04 &  & 2.28 & 1.33 & 0.12 \\
  1.74 & 0.74 & 0.04 &  & 2.29 & 1.00 & 0.10 \\
  1.75 & 0.60 & 0.06 &  & 2.31 & 0.83 & 0.08 \\
  1.76 & 0.58 & 0.05 &  & 2.32 & 0.58 & 0.06 \\
  1.77 & 0.43 & 0.05 &  & 2.34 & 0.67 & 0.08 \\
  1.79 & 0.37 & 0.05 &  & 2.36 & 0.60 & 0.07 \\
\enddata
\label{tab2}
\end{deluxetable}

\subsection{Photometry}

On August 7 2009 (UT), when OSIRIS was too warm to operate, an angular
differential imaging (ADI, Marois et al. 2006\nocite{Marois2006}) sequence was
taken with NIRC2 using four narrow-band filters in the $K$-band window (see
Table \ref{tab3}).  These filters were selected to sample, as
uniformly as possible, the full $K$-band region accessible with OSIRIS and thus
serve as a consistency check on the shape of the final, fully processed, OSIRIS
spectrum.  These narrow-band data are compared in Figure \ref{fig3} to
equivalent flux points obtained from the OSIRIS data convolved with the filter
response profiles.  The NIRC2 narrow-band colors are consistent with the OSIRIS
spectral shape on either side of the peak flux; however, HeIB (2.06 $\mu$m)
photometry is only consistent at $\sim$ 2-$\sigma$.  The overall agreement,
between these two non-contemporaneous data sets, processed with different
reduction software, provides some confidence that the planet's spectrum was
faithfully extracted from the speckle-contaminated data.

Also in 2009, new CH$_4$ (4\% S/L) photometry was obtained with the Near
Infrared Coronagraphic Imager (NICI; Toomey et al. 2003 \nocite{Toomey2003}) on
Gemini South.  The NICI CH$_4$ short/long filters more optimally probe the
strength of CH$_4$ absorption than the NIRC2 CH$_4$ short/long filters.  The
former are narrower, have less overlap, and have central wavelengths that
correspond well to the min and max fluxes seen in $H$-band spectra of mid to
late T dwarfs.  A comparison between these new photometric data and equivalent
band-integrated flux points from the OSIRIS spectrum is shown in Fig.
\ref{fig3}.  Here again the CH$_4$ short/long slope is in excellent
agreement with the OSIRIS spectrum.

Since the discovery of the HR8799 planetary system more broad-band observations
have been published and improvements have been made in the LOCI
\cite[]{Lafreniere2007} and ADI algorithms (at least those used by this group).
For the analysis presented here, the $H$-band photometry of \cite{Metchev2009}
has been adopted.  Also, $K_s$ data taken in 2010 have been analyzed and a new
$K_s$ magnitude (M$_{\rm Ks}$ = 14.15 $\pm 0.1$) has been obtained.  See Table
\ref{tab3} for a full list of the NIRC2 and NICI photometry used in this
study.

\begin{deluxetable}{lcccccccc}
\tablecolumns{6}
\tablewidth{120pc}
\tablecaption{NIRC2 and NICI Photometry}
\tablehead{
            \colhead{Filter} & \colhead{$\lambda_c$} & \colhead{Mag} &
\colhead{F$_{10 pc}$ (mJy)} & \colhead{ZP(Jy)} & \colhead{Ref} }
\startdata
              &              &              &                      &          &   \\
\multicolumn{2}{r}{NIRC2 broad-band}    &              &                      &           &   \\
\cline{1-2}
            J &      1.25    &     16.30    &      0.46 $\pm$ 0.07 &   1521.1  & Mar08 \\
            H &      1.63    &     15.08    &      0.94 $\pm$ 0.11 &   1010.0  & Met09 \\
        K$_s$ &      2.15    &     14.15    &      1.43 $\pm$ 0.11 &    654.2  & updated \\
 L$^{\prime}$ &      3.78    &     12.66    &      2.06 $\pm$ 0.21 &    238.5  & Mar08 \\
              &              &              &                      &           &   \\
\multicolumn{2}{r}{NICI narrow-band}    &              &                      &           &   \\
\cline{1-2}
CH4$_{short}$ &      1.59    &     15.18    &      0.88 $\pm$ 0.14 &   1035.7  & new  \\
CH4$_{long}$  &      1.68    &     14.89    &      1.06 $\pm$ 0.18 &    959.7  & new  \\
              &              &              &                      &           &   \\
\multicolumn{2}{r}{NIRC2 narrow-band}    &              &                      &           &   \\
\cline{1-2}
       He I B &      2.06    &     14.73    &      0.85 $\pm$ 0.07 &    665.9  & new  \\
    H2(v=1-0) &      2.13    &     14.16    &      1.52 $\pm$ 0.11 &    703.4  & new \\
Br$_{\gamma 2}$ &      2.16    &     13.93    &      1.66 $\pm$ 0.12 &    618.9 & new   \\
    H2(v=2-1) &      2.26    &     14.09    &      1.37 $\pm$ 0.11 &    593.5  & new \\
\enddata
\label{photo_tab}
\tablenotetext{}{
  Mar08  = Marois et al. (2008); 
  Met09  = Metchev et~al (2009)}
\end{deluxetable}

\subsection{Comparison to Kn3 spectrum}

On June 21st (UT) 2009, \cite{Bowler2010} observed \hrb\ using the OSIRIS
Kn3 narrow-band filter.  A comparison between the narrow and broad-band spectra
is shown in Fig. \ref{fig3}, and both spectra agree within the 1-$\sigma$
error-bars.  The Bolwer et al.  spectrum decreases in flux more noticeably than
the broad-band spectrum;  Bolwer et al.  attribute this decrease in flux to
possible weak CH$_4$ absorption.  The broad-band spectrum, which encompasses
much more of the CH$_4$ absorption band, indicates an even weaker CH$_4$
signature than indicated by the Kn3 spectrum.

\section{Empirical Comparisons}

The $H$ and $K$-band spectra of \hrb\ show several interesting features.  In
the $H$ band, a pronounced triangular shape indicative of weak collision
induced absorption (CIA) and low surface gravity is seen.  The spectrum also shows no
evidence of strong methane absorption (consistent with CH$_4$ on/off photometry)
as would otherwise be expected in cold T-type brown dwarfs.  At $K$ band, the
spectrum shows very deep water absorption bands.  As in the $H$ band, there is no
evidence of strong methane absorption or very strong CO absorption.  

\begin{figure}[!thb]
\plotone{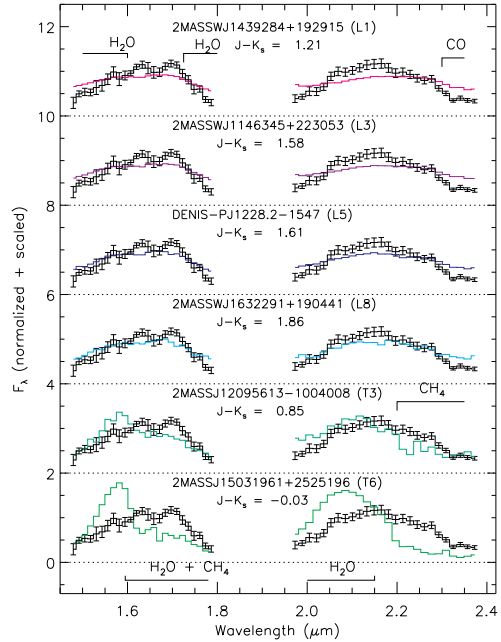}
\caption{OSIRIS $H$ anad $K$-band spectra of \hrb\ (plotted as F$_\lambda$)
compared to field brown dwarf spectra (taken from the SpeX spectral library).
Each band was normalized individually; the actual $J$-$K_s$ color is given
below each object name.  All of these field objects are bluer than \hrb.
\label{fig4}}
\end{figure}

\hrb\ occupies sparsely populated regions of near-IR color-magnitude
diagrams.  Despite this fact, it is still useful to compare the observed
spectra of \hrb\ to those of known brown dwarfs.  Figure \ref{fig4}
compares the $H$ and $K$-band spectra to a sequence of L and T dwarf spectra
from type T6 to L1.  The near-IR spectrum \hrb\ is fairly distinct from
these typical field brown dwarfs, which generally have weaker water absorption
and, for the T dwarfs, have deeper CH$_4$ absorption than seen in \hrb.
The fact that none of the hotter, L-type, brown dwarfs provide a good match to
\hrb\ is consistent with the low effective temperature deduced from cooling
track models \cite[]{Marois2008}.

Brown dwarf near-IR colors exhibit considerable spread within a given spectral
type and a number of peculiar (anomalously red or blue) L and T dwarfs are
turning up \cite[]{Kirkpatrick2008, Burningham2010} .  Consequently, while
standard BDs do not match \hrb\ very well, perhaps some of the more peculiar
BDs might.  The SpeX Prism spectral
library\footnote{http://web.mit.edu/ajb/www/browndwarfs/spexprism} provides
over 300 low-resolution L and T dwarf spectra from $\sim$ 0.65 to 2.55 $\mu$m
and includes a number of very red T dwarfs.  These spectra were rebinned to
match the resolution and sampling of the \hrb\ OSIRIS spectra and used to
find the best match to the $H$ and $K$ band spectra separately by minimizing
$\chi^2$.  All L and T dwarfs available in the SpeX library were used, except
known binaries.

\begin{figure}[t]
\plotone{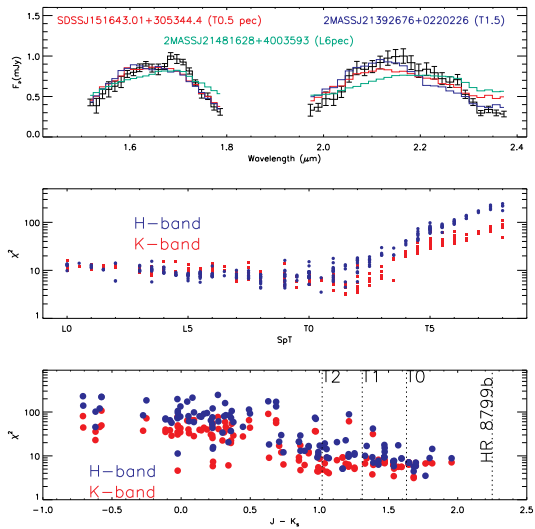}
\caption{
{\em top panel}: real brown dwarf spectra that best fit the $H$ and $K$ bands
of \hrb\ separately (red: best fit to $H$ band, blue: best fit to $K$ band). Each
spectrum is scaled arbitrarily to have the same total normalized flux.  Plotted
in green is the spectrum of 2MASS J2148+4003, the object found by
\cite{Bowler2010} to be the best overall match to \hrb.  {\em middle panel}:
$\chi^2$ values versus spectral type for real brown dwarfs compared to $H$
(blue) and $K$ (red) bands.  {\em bottom panel}: $\chi^2$ values versus $J-K_s$
color for real T-type brown dwarfs compared to the $H$ (blue) and $K$ (red)
bands. Vertical dotted lines indicate the $J-K_s$ color for \hrb\ as well as
typical colors for field T0, T1, and T2 dwarfs.
\label{fig5}}
\end{figure}

Figure \ref{fig5} shows the results of fitting the SpeX observations to
\hrb.  There is a clear $\chi^2$ minima between T0 and T2, with the best
fitting objects having spectral types T1.5 and T0.5 for the $H$ and $K$ bands,
respectively (within the range found by Bowler et al.\nocite{Bowler2010},
who fit their Kn3 spectrum to the same SpeX library).  An empirical fit to the
broad-band photometry was also performed by Bowler et al.; their best overall
match (a peculiar L6 2MASS 2148$+$4003) is compared to the $H$ and $K$-band
spectra in Fig. \ref{fig5}.  While this L6 is a good match to the broad
band photometry, it is not a good match to the broad-band near-IR spectral data.

It is interesting to note that, while fitting brown dwarf spectra is completely
independent of how well the broad-band colors agree (each band was normalized
individually for the $\chi^2$ calculation), there is a clear correlation
between $J$-$K_s$ color and $\chi^2$.  At first glance, this is simply
restating that the best fits are found among the late L and early T dwarfs,
which have the reddest $J$-$K_s$ colors.  However, in this case, the best
fitting objects are noticeably redder than the mean $J$-$K_s$ color for their
spectral type.  The lower panel of Fig. \ref{fig5} shows $\chi^2$ versus
$J$-$K_s$ color for all T dwarfs.  For both bands, the best fitting objects (a
T0.5 and a T1) are redder than the mean $J$-$K_s$ for T0 to T2 (vertical dotted
lines, values taken from Faherty et al.  2009\nocite{Faherty2009}).  While none
of the objects available in the SpeX library are as red as \hrb, there is a
clear indication that early T dwarfs, with anomalously red near-IR colors, are
the best match. This trend was also independently noted by Bowler et
al.\nocite{Bowler2010}. 

The best fit for $H$ band, SDSS J151643.01+305344.4, (here after SDSS 1516+30),
has $J$-$K_s$ = 1.77 and has been classified as T0.5 $\pm 1$ \cite[]{Chiu2006,
Burgasser2006} and  T1.5 $\pm 2$ \cite[]{Burgasser2010}.  While SDSS 1516+30
matches \hrb\ in $H$ band very well, the comparison is poor at $K$ band.
\cite{Stephens2009}, fitting their model atmospheres to available near-IR and
mid-IR spectra of SDSS 1516+30, found \teff\ = 1000 to 1100 K, $\log(g) = 4.5$
and a fairly cloudy atmosphere for such a low \teff\ ($f_{sed} =$
1 to 2, in the \cite{Ackerman2001} parlance).  This \teff\ range is lower than
typical for a T0.5, by 200 to 300K \cite[]{Stephens2009, Golimowski2004}.
\cite{Burgasser2010} identify SDSS 1516+30 as a weak, if not unlikely, binary
candidate; however, binarity  has yet to be tested with high resolution
imaging.  \cite{Leggett2007} have also noted this object's mid-IR colors are
best reproduced by models with thick clouds and stronger CO absorption than
predicted by pure chemical equilibrium, consistent with the high eddy diffusion
coefficient used by Stephens et al. in their model comparison.
 
The best fit for $K$ band is 2MASS J21392676+0220226 (here after 2M2139).  The
brown dwarf has $J$-$K_s$ = 1.68 and has been classified as T0
\cite[]{Reid2008}, T1.5 \cite[]{Burgasser2006}, and recently as T2.5 $\pm 1$
\cite[]{Burgasser2010}.  Not only does 2M2139 match \hrb\ reasonably well
in the $K$ band, the comparison is also about as good as SDSS 1516+30 in the
$H$ band.  \cite{Burgasser2010} suggest that this object is a strong binary
candidate, though no high-resolution imaging has been reported.  It is quite
possible that the object is single and simply non-standard with atmospheric
properties that are a mix of those found in the cloudy and cloud-free brown
dwarf photospheres.  The $K$-band spectrum of 2M2139 is similar to that of SDSS
J0758$+$42 (a T2), the brown dwarf found by Bowler et al. that best matched
their Kn3 spectrum.  The $H$-band spectrum of SDSS J0758$+$42, however, is
noticeably bluer than the \hrb\ OSIRIS $H$-band spectrum.

\begin{figure*}[t]
\plotone{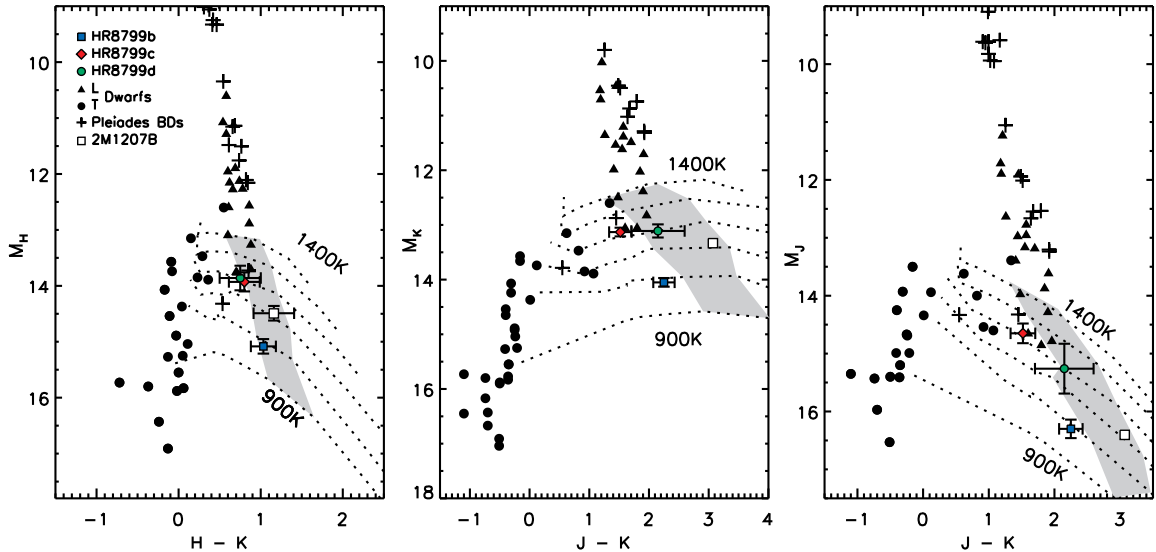}
\vspace{0.5cm}
\caption{Near-IR color-magnitude diagrams for brown dwarfs \cite[]{Leggett2002,
Knapp2004, Casewell2007} and the planets \twom\ and HR8799bcd.  Dotted lines
show color-magnitude tracks for chemical equilibrium models of constant T$_{\rm
eff}$ (900K to 1400K in steps of 100k, from top to bottom), $\log(g) = 4.0$,
and particle size (10 $\mu$m), but varying cloud thickness increasing from left
(no clouds) to right.  The shaded region indicates a zone of similar cloud
properties; namely where the exponential cloud weighting function begins at
similar heights (between $P_{min} = 1$ and 3 bar. In terms of cloud properties,
the young planet-mass objects appear to be an extension of the L-dwarf
sequence (see text for more details).
\label{fig6}}
\end{figure*}

In summary, the empirical comparisons suggest a spectral type of T1 $\pm 1$,
consistent with the T2 match that \cite{Bowler2010} found when comparing their
Kn3 spectrum to the SpeX library.  However, it appears as though only the
reddest T dwarfs in this spectral type range provide reasonable matches to the
\hrb\ spectra.  While a T1 is a close match, the near-IR colors of \hrb\ are
still well out of the normal range for this spectral type (being more
consistent with the mid L dwarfs; consequently, there are probably significant
limitations to what any near-IR spectral type can tell us about the basic
properties of \hrb.

\section{Model Atmospheres}

Recognizing that \hrb\ falls outside of the color-magnitude range of most known
substellar objects, comparing the observed data to synthetic spectra from model
atmospheres is the next logical step.   The {\tt PHOENIX} model atmosphere code
\cite[]{Hauschildt1997} has been used to explore a variety of physical
mechanisms that might explain the unusually red near-IR colors, low luminosity,
and lack of strong methane bands.  The version of {\tt PHOENIX} (v16) used here
has been substantially updated since the introduction of the {\tt PHOENIX}
``cond" and ``dusty" brown dwarf atmosphere models \cite[]{Allard2001}.   Some,
but not all, of the updates include the replacement of the original Allard
chemical equilibrium solver with a new, more robust and chemically complete,
equation of state\footnote{The new equation of state routine is known as the
Astrophysical Chemical Equilibrium Solver, or ACES (Barman \& Hauschildt, in
prep).}, a revised treatment of pure dust grains, improved alkali line profiles
\cite[]{Allard2007}, and new molecular line lists, most notably for methane
\cite[]{Warmbier2009, Hauschildt2009} and water \cite[]{Barber2006}.  Updates
to the cloud modeling and the addition of mixing-induced disequilibrium
chemistry are the most relevant here and are described in the following
sections.  
 
\subsection{A Simple Cloud Model}

Atmosphere models born from the study of brown dwarfs are commonly used to
estimate the properties of giant planets and to estimate detection yields for
direct imaging surveys, so it is natural to use such models as starting points
for an analysis of \hrb.  Arguably the most complex physical process to
include in substellar atmosphere models is the presence of clouds.  The term
cloud is used here to refer to solid or liquid particles suspended in the
atmosphere and, depending on the temperatures and pressures, can include a
complex mixture of species ranging from iron particles to ices.  The majority
of substellar atmosphere models are one-dimensional, assuming either
plane-parallel or spherical geometry, and consequently when clouds are included
their spatial properties can only be modeled radially.  Nevertheless, the
``patchiness" of clouds can be explored in an ad hoc fashion via a combination
of cloudy and cloud-free models \cite[]{Burgasser2002} or, more recently and
selfconsistently, within a single one-dimensional model \cite[]{Marley2010}.
For an extensive review and comparison of theoretical cloud modeling in
substellar atmospheric conditions, see \cite{Helling2008b}.

The observed luminosity and colors of \hrb\ have already been shown by
\cite{Marois2008} to be inconsistent with those of field brown dwarfs.
Therefore, by extension, these data are inconsistent with the predictions of
most published substellar atmosphere models, as they are often tuned to match
existing observations.  A large number of early brown dwarf models reproduced
the basic trends of L and T dwarfs by approximating the cloudy and cloud-free
conditions in a very simple manner; by simply turning on (cloudy) or off
(cloud-free) the opacity of dust in the atmosphere at its chemical equilibrium
temperature-pressure location \cite[]{Allard2001}.  This approach, taken to be
the extreme limits of cloud influence, is not expected to reproduce every
individual object well, especially those in the transition region from spectral
types L to T. In fact, most objects should fall between these two limiting
cases.  In order to match the photometric and spectroscopic observations of 
\hrb\ a different approach is required -- a cloud model that is intermediate
to these two limits is needed.
 
Recent improvements in the atmospheric modeling of transition brown dwarfs have
been made, with most efforts attacking the cloud problem by adding one or more
new parameters that regulate the cloud thickness, particle density, grain size
distribution and so forth \cite[]{Ackerman2001, Marley2002, Tsuji2005,
Cooper2003, Burrows2006}.  However, see \cite{Helling2008a} for a completely
different approach.  The purpose of this paper is not to develop a new complex
cloud model, but rather to identify the major physical processes that shape the
spectral properties of \hrb.  To that end, a simple parameterized cloud model
is adopted for the analysis presented here.   The lower boundary (cloud base)
is established by the intersection of the temperature-pressure (T-P) profile and
the relevant chemical equilibrium condensation curve ($P_c$, using pressure as
a proxy for height). The behavior of the cloud above $P_c$ is determined by a
single free parameter ($P_{min}$).  The equilibrium dust concentration is
multiplied by a function that is 1 for $P_{gas} \ge P_{min}$ and decays
exponentially for $P_{gas} < P_{min}$.  If $P_{min} > P_c$, then the maximum
dusty-to-gas ratio is also lowered relative to an equilibrium cloud.
Examples of vertical cloud properties are discussed below in more detail. The
particle sizes follow a log-normal distribution with a prescribed modal size
(\azero) from 1 to 100 $\mu$m that is independent of height.  This cloud model
shares some similarities with other parameterized cloud models
\cite[]{Tsuji2005, Burrows2006}; however, rather than accounting for a small
representative set of grains, all thermodynamically allowed grains with
opacities in the \phoenix\ databases are included in the total cloud opacity;
see \cite{Ferguson2005} for a list of grains included in the model.

Figure \ref{fig6} shows the location of \hrb\ relative to L and T brown
dwarfs in three near-IR color-magnitude diagrams.  Synthetic color-magnitude
sequences are also shown for chemical equilibrium models (K$_{zz}$ = 0) with
radius set to that of Jupiter.  Each sequence corresponds to a different \teff\
(with \logg\ = 4.0) and different cloud thickness varying from no clouds (at
the blue end) to maximum cloud thickness (at the red end).   These sequences of
models are not associated with any evolutionary cooling tracks since the radius
has been arbitrarily set to that of Jupiter.  Such sequences illustrate the
color trends produced by changing vertical cloud thickness (note this is
different from the uniform sinking of clouds).  Figure \ref{fig6} shows that
the slope in a CMD across the L-to-T transition region, including the $J$-band
brightening, is reproduced very well by atmospheres that have clouds smoothly
decreasing in vertical thickness.  However, as pointed out by
\cite{Burrows2006}, reproducing color-trends does not guarantee good
reproduction of near-IR spectra; the models shown in Fig. \ref{fig6} have not
been compared to spectra of field brown dwarfs (an exercise left to another
paper).

Figure \ref{fig6} also shows that very red objects with relatively low
luminosities, like \hrb, can be reproduced by allowing low \teff\ models
(typically assumed to be in the regime of cloud-free photospheres) to have
clouds extending across their photosphere.  The location of \hrb, in all three
CMDs is intersected by a model with \teff\ = 1000K, \logg\ = 4.0, and a cloud
that starts to fall off near 1 bar of pressure (though the precise $P_{min}$
matching the position of \hrb\ is slightly different in each CMD).  Figure
\ref{fig6} also suggests that the cloud properties of the HR8799 planets and
2m1207b are similar to the those of late-type L dwarfs.  The comparison in Fig.
\ref{fig6} does not directly identify a best fit to the ensemble of data, but
simply illustrates the impact that cloud thickness has on the near-IR colors of
an object with typical substellar temperatures and gravities and points to a
reasonable place to start looking for a good match to \hrb, namely \teff $\sim
1000$K and $\log(g) \sim 4.0$.

\subsection{Disequilibrium Chemistry and Vertical Mixing}

The majority of atmosphere models for brown dwarfs and exoplanets assume local
chemical equilibrium; in other words, the mole fractions for each chemical
compound is determined by the local temperature and pressure, and overall
element abundances.  For most atmospheric compounds, this is an excellent
approximation as the chemical reactions that establish the equilibrium mole
fractions are rapid compared to the mixing time scales (even in the convection
zone).  However, this is not always the case and it has been shown that, for
example, the reactions that ultimately replace CO with CH$_4$ can be very slow
in Jovian and brown dwarf photospheres.  The same can be true for N$_2$ and
NH$_3$.  \cite{Prinn1977} showed that vertical mixing of CO can be important in
Jupiter and, later, \cite{Fegley1996} predicted that this would also be the
case for brown dwarfs.  The spectroscopic detections of CO in several cool
T-type brown dwarfs \cite[]{Noll1997, Saumon2000, Saumon2006, Geballe2009} along with
weaker than predicted fluxes from $\sim 4.5$ $\mu$m ground and space-based
photometry \cite[]{Patten2006, Leggett2007} have established disequilibrium CO
mole fractions as a common occurrence in brown dwarfs.  Non-equilibrium
chemistry is, therefore, a plausible mechanism for the apparently weak CH$_4$
absorption in \hrb.  

Disequilibrium CO/CH$_4$ and N$_2$/NH$_3$ chemistry has been added to \phoenix\
following \cite{Smith1998}.  The reaction time scales (in seconds) for
$N_2 \rightarrow NH_3$ and $CO \rightarrow CH_4$ are from \cite{Lodders2002}
and \cite{Yung1988}, respectively:  

\begin{equation}
\tau_{N_2} = \frac{1}{\kappa_{N_2} N(H_2)}
\end{equation}

\begin{equation}
\kappa_{N_2} = 8.45 \times 10^{-8} exp \left(\frac{-81515}{T}\right)
\end{equation}

\begin{equation}
\tau_{CO} = \frac{N(CO)}{\kappa_{7} N(H) N(H_2CO)}
\end{equation}

\noindent where $T$ is the gas temperature in Kelvin, $\kappa$ are the rate
coefficients in $cm^3/sec$ (with $\kappa_{7}$ interpolated from tabulated
values in Yung et al. 1988\nocite{Yung1988}) and N(species) is the number density of the
specified species in $cm^{-3}$.

\begin{figure}[!tb]
\plotone{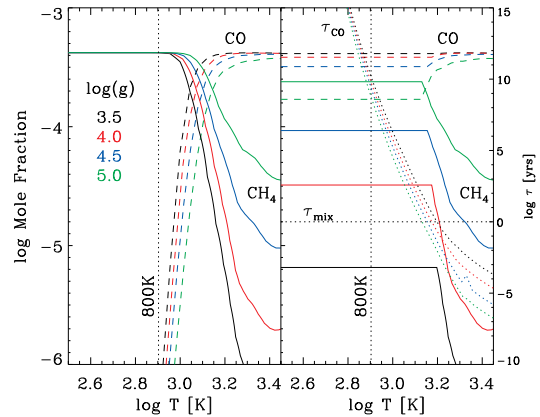}
\caption{
CO (dashed lines) and CH$_4$ (dotted lines) mole fractions for a cloudy, solar
metallicity, model atmospheres with \teff\ = 800K and four different surface
gravities (corresponding colors indicated in the figure legend).  Chemical
equilibrium mole fractions are plotted in the left panel.  Non-equilibrium mole
fractions are plotted in the right panel, where the mole fractions are quenched
at different depths determined by the intersection of the CO chemical reaction
and mixing time scales ($\tau_{CO}$ and $\tau_{mix}$, dotted lines).  The latter
time scale was arbitrarily set to 1 yr.  In both panels, the vertical dotted
line marks the approximate location of the photosphere (taken here to be where
the local temperature equals the effective temperature).
\label{fig7}}
\end{figure}

The mixing time scale ($\tau_{mix}$) is computed using mixing length theory in
the convection zone, and set equal to $L_{\rm eff}^2 / K_{zz}$ in the radiative
zone, where the effective length scale ($L_{\rm eff}$) is determined following
the recipe in \cite{Smith1998}.  For the range of models explored here, $L_{\rm
eff}$ was found to be  $\sim 0.1$ to 0.2$H_{P}$, with $H_{P}$ equal to the
pressure scale height. This range of $L_{\rm eff}$ is comparable to that found
for Jupiter \cite[]{Smith1998, Visscher2010}.  The major unknown when modeling
vertical mixing is the eddy diffusion coefficient ($K_{zz}$) which, as is
frequently done, is assumed here to be an adjustable parameter typically
ranging from 100 to $10^8\ cm^2\ sec^{-1}$.  With these chemical and mixing
time scales, the mole fractions of CO/CH$_4$ and N$_2$/NH$_3$ are ``quenched"
at a specific depth in the atmosphere where $\tau_{mix} = \tau_{chem}$.  Below
this point in the atmosphere, where $\tau_{mix} > \tau_{chem}$, chemical
equilibrium prevails.  This simple quenching model has been used in several
recent studies of brown dwarfs and giant planets \cite[]{Hubeny2007, Fortney2008}.

The mixing and chemical time scales alone do not determine the final quenched
mole fractions, the atmosphere temperatures and pressures still establish these
values at the quenching depth.  Consequently, effective temperature, surface
gravity, and metallicity play an important role.  Of particular importance is
the relative location in the atmosphere of equal CO/CH$_4$ equilibrium mole
fractions ($P_{CO = CH_4}$) and where $\tau_{mix} = \tau_{chem}$ ($P_{q}$).  If
$P_{q} > P_{CO = CH_4}$, the CO/CH$_4$ ratio can be greater than 1 in cool
substellar atmospheres where CO would otherwise be predicted to be orders of
magnitude less than CH$_4$.

As shown in Figure \ref{fig7}, the magnitude of the disequilibrium effect
is sensitive to surface gravity, as this quantity acts to shift the locations
of $P_{q}$ and $P_{CO = CH_4}$. As gravity decreases, the crossing point of the
radiative and mixing time scales moves to deeper layers (higher T), while the
crossing of the equilibrium mole fractions of CO and CH$_4$ moves outward.  For
$K_{zz}$ values typically used in brown dwarf models and at low surface
gravities expected for young exoplanets, the CO/CH$_4$ ratio can become
significantly greater than 1 -- indicating a role-reversal in the dominant
carbon-bearing species at photospheric depths.  Such a role reversal has not
been explored thoroughly in brown dwarfs as their surface gravities and
effective temperatures almost always place the time scale crossing point at a
pressure lower than that of the equilibrium CO/CH$_4$ crossing point, resulting
in a CO/CH$_4$ ratio always less than 1 (though still greater than equilibrium
chemistry predictions).  \cite{Hubeny2007} also noted a gravity dependence for
non-equilibrium chemistry, but did not explore gravities below \logg = 4.5 nor
encountered a CO/CH$_4$ reversal.  Since a low surface gravity and effective
temperature are possible outcomes for \hrb, a lack of (or very weak) CH$_4$
could be caused by disequilibrium chemistry.
 
\subsection{Photochemistry}

\hrb\ is located a mere 68 AU from its host star and receives far too
little stellar flux for irradiation to have a significant influence on its
atmospheric thermal structure (at near-IR photospheric depths).  However,
HR8799A is much more luminous ($\sim 5$ L$_{\odot}$) and hotter (\teff = 7500K) than
the Sun and the UV flux from HR8799A may be sufficient to produce
photochemical reactions in the atmosphere of \hrb\ and certainly for the inner
three planets.  Based on a synthetic spectrum from a stellar model atmosphere
calculation tailored for HR8799A (also using \phoenix, with atoms and ions
treated in non-local thermodynamic equilibrium), \hrb\ is estimated to receive
$\sim$ 8 times the flux that Jupiter receives from the Sun for $\lambda <
2000$\AA, and about 10 times as much flux as Neptune for $\lambda < 3000$\AA.
Given the larger atmospheric temperatures and possibly extreme non-equilibrium
CO/CH$_4$ mole fractions compared to the giant planets in the Solar System, the
photochemical reactions in \hrb\ will also be quite different from those found
in the Solar System.  Consequently, photochemistry could impact the model
comparisons discussed below.  Unfortunately, photochemical reactions are not
included in the version of \phoenix\ used here. Future observations and
modeling will be needed to address this possibility.

\section{Model Comparisons}

The complete set of \hrb\ broad-band photometry covers a broader wavelength
range than the OSIRIS spectroscopic observations and, therefore, are likely to
provide the greatest leverage when estimating effective temperature and,
potentially when estimating the cloud properties.  \cite{Marois2008} did not
perform a model fit to the discovery photometry but, rather, focused on the
parameters inferred from theoretical cooling tracks and the observed luminosity
and age range.  With new and improved photometry available, new model fits to
these data are now warranted (see \cite{Bowler2010} and \cite{Currie2011} for
independent model comparisons to similar photometric data and different model
atmospheres).  Additionally, the near-IR spectra are likely to contain the
greatest information on surface gravity, chemical composition (via broad-band
absorption features), and also on the cloud properties.  The goal in this
section is to identify a single model atmosphere that best fits all three
observational data sets and to understand what physical mechanisms play
dominant roles in shaping the overall spectral energy distribution.

\begin{figure}[!tb]
\plotone{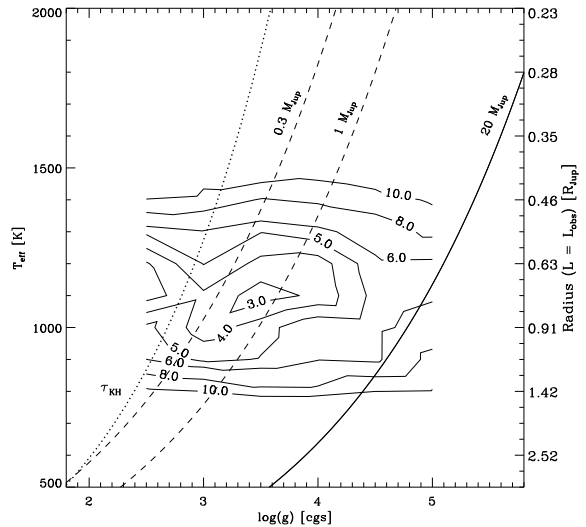}
\caption{
Mean $\chi^2$ contours obtained by comparing model spectra to the broad-band
photometry (Table 2) and OSIRIS $H$ and $K$ spectra (Table 3).  The
best fitting model has \teff = 1100K and \logg = 3.5.  The right ordinate
indicates the radii required to match the observed luminosity of \hrb\
given the effective temperatures on the left ordinate. Using these radii and
the surface gravities along the abscissa, lines of constant mass can be plotted
(dashed lines for 0.3 and 1 \mjup, solid line for 20 \mjup). These lines of
constant mass are {\em independent} of atmosphere or evolution model
predictions.  Points along the dotted line are consistent with a
Kelvin-Helmholtz cooling time of 1 Myr.
\label{fig8}}
\end{figure}

\subsection{Solar Abundance Model Comparison}

The cloud and disequilibrium chemistry models described above were incorporated
into \phoenix.  A modest grid of synthetic spectra was computed with five key
parameters considered: \teff\ [= 800K -- 1500K, 100K increments], \logg\ [= 2.5
-- 5.0, in 0.5 increments and CGS units], \kzz\ = 10$^4$ cm$^{2}$ s$^{-1}$,
\pmin\ [= 10$^5$, 10$^6$, 2$\times 10^6$, 4$\times 10^6$, 6$\times 10^6$,
8$\times 10^6$, 1$\times 10^7$ dynes cm$^{-2}$], \azero\ [ = 1, 5, 10, 100
$\mu$m]. Solar abundances were adopted for this grid \cite[]{Asplund2005}, but
the question of non-solar metallicity will be discussed below. 

Given the large number of free parameters and the sparse sampling of the model
grid, obtaining a traditional fit by $\chi^2$ minimization that samples a
continuous distribution of parameters is beyond the scope of this paper.
Instead, a $\chi^2$ was computed for each model spectrum from the grid compared
to each of the three data sets (photometry, $H$ and $K$-band spectra).  When
computing a $\chi^2$ for the photometric comparison, each model spectrum was
convolved with the appropriate filter response functions to produce synthetic
photometry in $J$, $H$, $K_s$, and $L^\prime$ bands along with the filters used
by \cite{Hinz2010} and \cite{Currie2011}; Mbar, $z$, and a filter centered at 3.3
$\mu$m.  When comparing models to the OSIRIS spectra, the synthetic spectra
were first smoothed (with a Gaussian kernel) to the native instrumental
response for OSIRIS ($R \sim 4000$), then binned down to the same wavelength
sampling (and in the same manner) as done for the OSIRIS data cubes.  The best
fitting model was taken to be the model with smallest mean  $\chi^2$ for the
three data sets.  Figure \ref{fig8} shows the mean reduced-$\chi^2$
distribution across \teff\ and \logg\ for the model comparisons to all three
sets of data.  The best fit is located at \teff = 1100K and \logg = 3.5.
Formal error-bars for the fit have not been determined; however, models within
the mean $\chi^2$=4 contour are all reasonable fits to the data.  Therefore,
the error on the fit is close to $\pm$100K for effective temperature and
$\pm$0.5 dec for surface gravity, and comparable to most brown dwarf model fit
uncertainties found in the literature. 

\begin{figure}[!tb]
\plotone{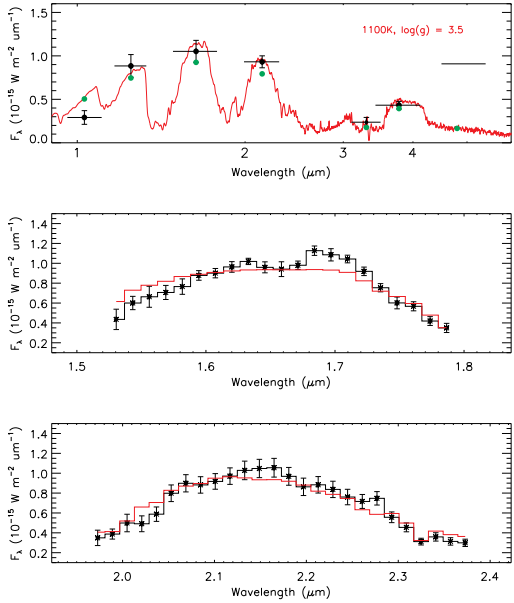}
\caption{
Comparison between the best fitting solar-abundance model and the broad-band
photometry (top panel) and OSIRIS broad-band $H$ and $K$ spectra (middle and
bottom panels).  Observations are show in black with 1-$\sigma$ error bars, the
model spectrum in red and model photometry in green.  The photometric upper
limit from \cite{Hinz2010} is plotted as a horizontal bar.
\label{fig9}}
\end{figure}

Figure \ref{fig9} compares the best fit to each of the three data sets.  The
model spectrum agrees very well with the photometry. The differences seen at
$J$, $H$, $K_s$, and $L^\prime$ are within the 1-$\sigma$ error bars.  The
model only slightly under predicts the flux at 3.3 $\mu$m, compared to the
\cite{Currie2011} detection.  The flux measurement at 3.3 $\mu$m excludes a
large number of models, especially at high \teff, and is quite sensitive to the
presence or absence of CH$_4$. Observations of other systems (e.g., \twom)
in this band-pass would be very useful.

At $H$ band, the model spectrum is marginally consistent with the observed
spectrum. The slope at the red and blue ends are not as steep as observed and,
overall, the model spectrum is too flat.  It is possible that the residual flux
from the bright speckle feature seen near $\sim$ 1.7$\mu$m (see Fig.
\ref{fig3}) remains in the final observed spectrum and is responsible for
the larger disagreement seen at this wavelength.  As for the red and blue
$H$-band slopes, these wavelength regions have historically been difficult to
fit mainly due to missing CH$_4$ line opacity. Consequently, the level of
disagreement at $H$-band is not too surprising.  At $K$ band, the model
compares favorably to the observed spectrum.  The red and blue slopes are in
good agreement as is the location of the peak flux. 
 
When the data sets are compared to the models individually, slightly different
results are obtained.  The best fitting model to the photometry has \teff\ =
1000K and $\log(g) = 5.0$.  $H$ band is best fit by a model with \teff\ = 1200K
and $\log(g) = 3.0$ (clearly favoring low gravity); however, this model has
colors that are too blue when compared to observed photometry, especially in
the $z$-band.  At $K$ band, the best fit has  \teff\ = 1100K and $\log(g) =
3.0$.  When the mean $\chi^2$ for $H$ and $K$ is minimized, the best fit is
\teff\ = 1200K and $\log(g) = 3.5$.  Each of these best fit models has a
slightly different cloud thickness, but all have \kzz\ = 10$^4$ cm$^{2}$
s$^{-1}$ and \azero\ = 5 $\mu$m. 

\begin{figure}[!tb]
\plotone{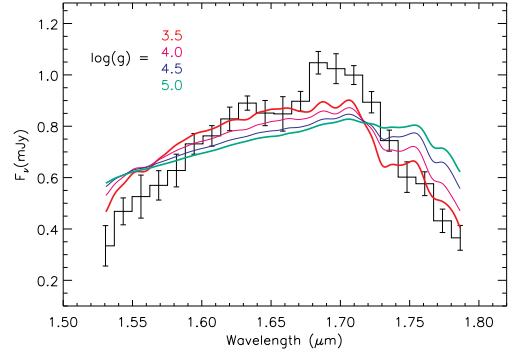}
\caption{
Model gravity sequence compared to the OSIRIS $H$-band spectrum.
All model parameters are identical the best fitting model shown
in Fig. \ref{fig9} except for gravity, which varies from $\log(g) =$
3.5 to 5.0 (cgs units).  The triangular shape of the $H$-band is
best explained by low surface gravity.
\label{fig10}}
\end{figure}

\subsection{Inferred Atmospheric Properties}

Despite the imperfect fit to the $H$-band, there is strong evidence for low
surface gravity.  In Figure \ref{fig10} the $H$-band OSIRIS spectrum is
compared to a sequence of model spectra with different gravities but with all
other parameters equal to those of the overall best fit shown in Fig.
\ref{fig9}.  The observed triangular shaped pseudo-continuum is best
reproduced by the lowest-gravity model.  The shape of the $H$ band is often used
as a gravity indicator in L-type brown dwarfs, where low gravity objects have
distinct triangular spectra \cite[]{Lucas2001, Kirkpatrick2006} compared to the more
rounded spectra of older, higher gravity, field brown dwarfs.  This
gravity-dependence of the $H$-band is produced by weakening of the continuous
opacity produced by collision-induced absorption (CIA) as gravity decreases,
relative to the molecular line opacity (most significantly relative to that of
H$_2$O).  When CIA is weak, the $H$-band shape is defined by the natural
triangular shape of the H$_2$O bands on either side of $\sim 1.7 \mu$m.
However, when the surface gravity is high, the larger pressures increase the
CIA to a point where it can become the dominate opacity source between the two
H$_2$O bands on either side of the peak.  The smooth wavelength dependence of
CIA rounds the $H$-band peak \cite[]{Borysow1997, Kirkpatrick2006, Rice2011}.
For field brown dwarfs, the effectiveness of $H$-band as gravity indicator is
decreased for field T dwarfs as CH$_4$ opacity dominates the near-IR and
competes with that of H$_2$O around the $H$ band.  However, if CH$_4$ mole
fractions are reduced by vertical mixing and non-equilibrium chemistry at low
gravity and low \teff, as is likely the case for \hrb, then the behavior of the
$H$ band with gravity seen in hotter L-dwarfs will persist even down to
effective temperatures most often associated with T-dwarfs.  

\begin{figure}[!tb]
\plotone{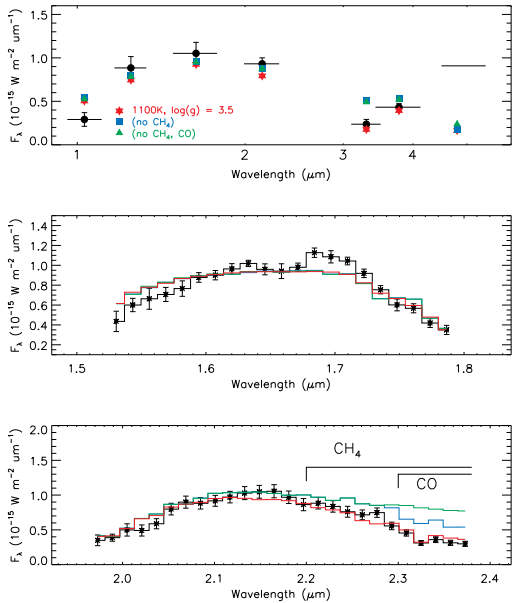}
\caption{
Same, best fitting model as shown in Fig. \ref{fig9} (red), compared to
models with the same parameters and temperature-pressure profile, but with
CH$_4$ opacities removed (blue) or both CO and CH$_4$ opacities removed
(green).
\label{fig11}}
\end{figure}

Water is by far the main molecular opacity source shaping the OSIRIS spectra.
The combination of high \kzz\ and low \logg\ leads to, in the models used here,
a sharp reduction in the CH$_4$ mixing ratio (e.g.  see Fig. \ref{fig7})
with a correspondingly large increase in the CO mixing ratio.  Figure
\ref{fig11} compares the best fit model to those with the same parameters and
temperature-pressure structure, but with CO or CH$_4$ opacities removed.  Weak
CH$_4$ absorption is evident in the $K$ band, as is a possible contribution
from CO; however, the strongest evidence for CH$_4$ absorption is the
\cite{Currie2011} detection (and \cite{Hinz2010} upper limit) at 3.3 $\mu$m
where the flux is clearly suppressed by CH$_4$. Unfortunately, the
\cite{Hinz2010} upper limit near 4.5 $\mu$m is not sensitive enough to
distinguish between the majority of the models explored here and, therefore,
does not provide strong constraints on CO.  

\begin{figure}[!tb]
\plotone{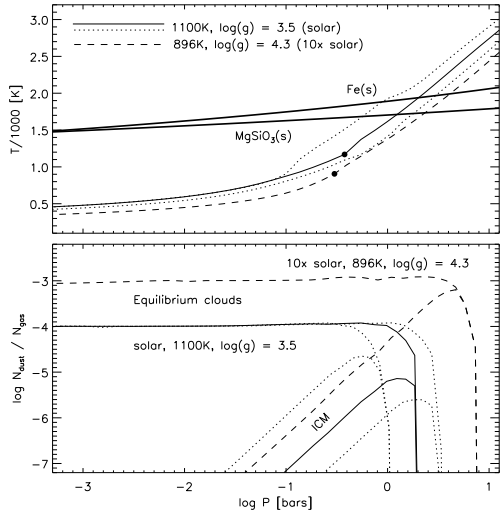}
\caption{
{\em Top} : Temperature versus pressure for the best fitting model identified
in Fig. \ref{fig8} (solid line) and for two models with the same \teff,
\logg\ and \kzz, but with different cloud model parameters (dotted lines).
Also shown (dashed line) is the best fitting metal rich model (Fig. \ref{fig14}).
Condensation curves (thick solid lines) for Fe and MgSiO$_3$ are plotted to
indicate the general location of the cloud base, located at the intersection of
these curves.  Other grains (e.g. Mg$_2$SiO$_4$, Al$_2$O$_3$, Ni, etc.) have
condensation curves in same general location. {\em Bottom} : Dust fraction
versus pressure for the same models shown in the top panel (in order from left
to right).  A constant dust fraction with decreasing pressure is a hallmark of
the pure chemical equilibrium cloud (e.g. the 'DUSTY' models of Allard et al.
2001).  In contrast, the intermediate cloud model (ICM) used here can have
lower peak dust fraction than the equilibrium cloud and decays with height (see
text for details).
\label{fig12}}
\end{figure}

The best fitting, solar abundance, atmosphere model for \hrb\ (shown in
Fig.  \ref{fig9}) has a cloud base just below $P = 1$ bar with a composition
that is typical for brown dwarf models with \teff\ $> 1000$K.
Magnesium-silicate (e.g. forsterite) and iron grains are the most abundant,
with a mixture of other grains with lower mole fractions.  The structure of the
cloud has many characteristics in common with brown dwarf cloud predictions,
namely an abrupt drop in the gas-to-dust ratio at higher pressures and
temperatures, where condensation is no longer thermodynamically favored. This
cloud base is determined by the intersection of the T-P profile and the
relevant condensation curves (see Fig. \ref{fig12}).  At lower temperatures
and pressures, the gas-to-dust ratio drops exponentially (by design in this
study, see lower panel of Fig. \ref{fig12}) in a similar manner as the
smooth decrease in dust predicted by the models of Helling \& Woitke and Marley
\& Ackerman (see \cite{Helling2008b} for a comparison).  Location and vertical
extent are the most important aspects of the cloud structure for \hrb\
preferred by this work.  Generally, the cloud base retreats to larger pressures
with decreasing \teff\ and fixed gravity; however, the low gravity of the best
fitting \hrb\ model counteracts this trend.  As gravity decreases the T-P
profile shifts to lower pressure, moving the cloud base along with it.  At
\logg\ $\sim 3.5$, the cloud base sits near the photosphere with vertical
extent that is far less than an equilibrium cloud model (see Fig.  \ref{fig12}),
but sufficient to produce redder colors than cloud-free brown dwarf models with
the same \teff\ and \logg.  

The modal particle size for \hrb, favored by this work is around 5 to 10
$\mu$m. Grains of this size are within the range predicted by the Marley \&
Ackerman and Helling \& Woitke for models with \teff\ $\sim$ 1000K. Both groups
predict sizes between 1 and 100 $\mu$m at high pressures ($>$ 10 bar) and sizes
that decrease below a micron at low pressures ($<$ 1 bar).   \cite{Burrows2006}
chose \azero = 100 $\mu$m (at the high end of most predictions) for their set
of reference model atmospheres arguing that, in general, L-type brown dwarfs
are best fit when larger grain sizes are assumed but acknowledged that such
large grains make fitting the reddest late L dwarfs difficult.  As already
discussed above, the observed near-IR spectra of the reddest late L and early T
dwarfs come closest to matching the $H$ and $K$-band spectra of \hrb\ and,
thus, perhaps it should not be surprising that the best model fits are found
among those with smaller grains.  However, the simulations of \cite{Cooper2003}
predict that particle size should actually increase with decreasing gravity.
When comparing grain sizes adopted by different models, it is important to
remember that grain size alone does not determine the total cloud opacity --
composition also plays an important role.  In the cloud model used here a
mixture of grains are included in the total opacity and such a cloud can be
more opaque than a cloud including opacity from only pure silicate grains of
the same size.
 
\section{Discrepancy Between Atmosphere and Evolutionary Models}

Since the distance to HR8799 has been precisely measured (39.4 $\pm$ 1.0 pc,
van Leeuwen 2007\nocite{vanLeeuwen2007}), the luminosity of \hrb\ is known
($\log L_{\rm obs}/L_{\odot} = -5.1$) and, therefore, radii can be assigned to
each \teff\ value on the left vertical axis of Fig.  \ref{fig8}; these radii
are given by the right, non-linear, vertical axis.  Using these radii and the
$\log(g)$ axis values, lines of constant mass can be plotted that are
completely independent of interior cooling track models.  Dynamical stability
arguments suggest that, in order for the three planets to be as massive as
indicated by the cooling tracks, they must be in double 2:1 resonances and have
masses $\le 20$ \mjup\ \cite[]{Fabrycky2010}.  Using 20\mjup\ as an upper limit
on the mass of \hrb\ sets a \teff /\logg\ boundary in Fig. \ref{fig8}
(right-most solid line).  This limit was the motivation for not extending the
model grid to \logg\ greater than 5.0.  A second boundary, at lower surface
gravity, can be set using the Kelvin-Helmholtz time scale ($\tau_{\rm KH}$),
which reflects the approximate scaling of detailed planetary cooling curves.
Since the planets almost certainly are older than 1 Myr, $\tau_{\rm KH}$ = 1
Myr is a very conservative limit and is indicated by the left-most dotted line
in Fig.  \ref{fig8}.  In other words, planets located to the left of this
dotted line could not maintain the observed luminosity of \hrb\ for more
than 1 Myr (under the usual assumption that the primary energy source is
gravitational potential energy).  This time scale limit was used to exclude a
second minimum at \logg\ = 2.5 near \teff\ = 1100 and 1200K.  For \teff\ =
1100K $\pm$100K, the radius must be 0.75 \rjup $^{+0.17}_{-0.12}$ to match
\lobs.  Using this radius and the best fit surface gravity, $\log(g)$ = 3.5
$\pm$ 0.5, the corresponding mass is just under 1 Jupiter-mass ($\sim$ 0.72
\mjup $^{+2.6}_{-0.6}$). a factor of $\sim$ 10 lower than predicted by
evolutionary cooling tracks.  A radius of 0.75 \rjup\ is also $\sim 30$\%
smaller than cooling track predictions.

\begin{figure}[!tb]
\plotone{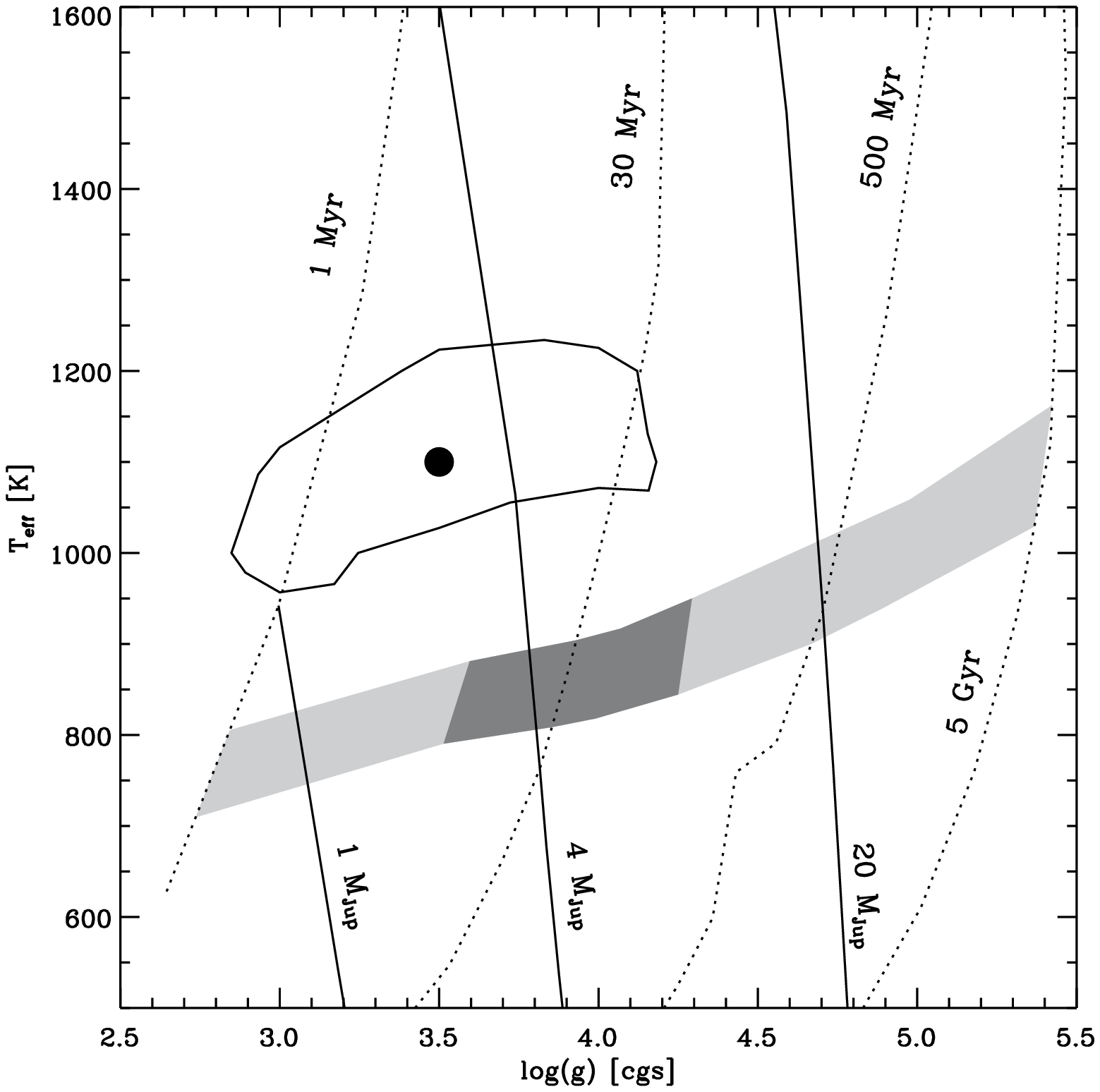}
\caption{
Plotted are lines of constant mass (solid) and isochrones (dotted) from
\cite{Baraffe2003} cond evolution models.  The shaded region covers the range of
\teff\ and \logg\ predicted by these evolution models for an object with the
same luminosity as \hrb\ (the darker region brackets 10 and 120 Myrs).  The
best fitting \teff\ and \logg, based on fitting solar abundance atmosphere
models (see text for details), is indicted by the filled symbol surrounded by
the $\chi^2 = 4$ contour from Fig. \ref{fig8}.
\label{fig13}}
\end{figure}

While the implied mass of $\sim 0.7$ \mjup\ is comfortably within the ranged
needed for dynamical stability, it is difficult to form objects, either by
core-accretion or by gravitational instability, with this mass, radius ($\sim
0.75$ \rjup), and the planet's observed luminosity.  This combination is
inconsistent with our understanding of degenerate or partially-degenerate H+He
gaseous bodies with any reasonable heavy-metal composition or of any age
\cite[]{Baraffe2008}.   Figure \ref{fig13} compares the \teff\ and \logg\
values from the atmosphere model fit to those predicted by evolution models
from \cite{Baraffe2003}.  While the best fit surface gravity is within the
expected range, an effective temperature of 1100K is $\sim 300$K above the
temperatures predicted by all current hot-star cooling tracks, for an age less
than 100 Myrs.  The high effective temperature would be consistent with
evolution predictions if the system was old ($> 1$ Gyr) and \logg\  was greater
than $\sim 5$.  However, as shown above, such a high surface gravity would be
very difficult to reconcile with the $H$-band spectrum and would have serious
dynamical stability implications \cite[]{Fabrycky2010, Marois2010, Currie2011}.  

\begin{figure}[!tb]
\plotone{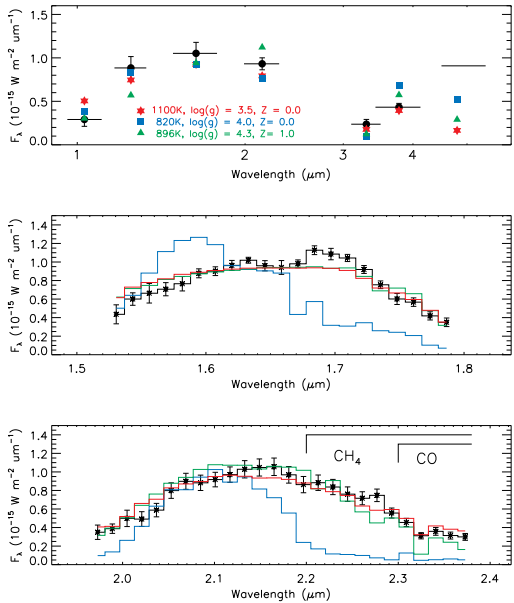}
\caption{
Same, best fitting, model as shown in Fig. \ref{fig9} (red), compared to the
best fitting model that has \teff\ and \logg\ consistent with standard ``hot
start" evolution models (green).  Also plotted are the synthetic spectra and
photometry for the chemical equilibrium atmosphere model (blue) from
\cite{Marois2008}. See text for details.
\label{fig14}}
\end{figure}

Giant planet evolution models are known to become increasingly unreliable at
young ages for the simple reason that planet formation is not well understood
and, thus, the initial conditions are not known.  This fact was highlighted by
\cite{Baraffe2002}, with a focus on the formation of brown dwarfs and low mass
stars from the collapse of molecular clouds (the main realm of applicability
for the common cond and dusty cooling tracks).  More recently,
\cite{Marley2007} revisited the question of initial conditions by comparing the
evolution of planets starting with high \teff\ and luminosity (``hot start", as
often adopted for brown dwarfs and motivated by the Stahler birth line), to the
evolution of planets starting with the colder and less luminous initial
conditions (``cold start") that emerge from core-accretion planet formation
models.  Marley et al. concluded that young planets that formed by
core-accretion are cooler, smaller, and thus fainter than planets formed by
single-mode gravitational collapse within a disk or giant molecular cloud.
Given that \hrb\ is clearly underluminous compared to the hot-start models
and overluminous for the cold-start models \cite[]{Marley2007, Fortney2008}, it
is unlikely that initial conditions alone are responsible for the offset
between the atmosphere-inferred properties and the evolution track predictions
(a situation similar to that of \twom, as discussed below).

A $\sim 300$K offset in effective temperature is similar in magnitude to
offsets in effective temperature found for a number of M, L and T-type binary
brown dwarfs \cite[]{Liu2008,Liu2010,Konopacky2010,Dupuy2010}.  This suggests
that a systematic offset may exist between the atmosphere and interior model
predictions.  A 10 to 20 \% difference in \teff\ is comparable to the
differences between cloudy and cloud-free tracks in the 10 and 100 Myr range
and 10 \mjup \cite[]{Chabrier2000}, so it is still unclear what fraction of
such an offset could be resolved by improvements purely in the modeling of the
interior or the atmosphere.  However, the increasing number of free parameters
in substellar atmosphere models is a clear indication that they are even more
complex than previously imagined.  Thus, the possibility still remains that
hidden among the free parameters is an atmosphere model with \teff\ and \logg\
that matches the evolution model predictions.  

\subsection{Non-solar Abundance Model Comparisons}

The need for an intermediate cloud model (constrained vertically), to match the
observed broad-band photometry of all three HR8799 planets, was first discussed
by \cite{Marois2008}.  \cite{Marois2008} showed that a synthetic spectrum from
a solar abundance model atmosphere with an intermediately thick cloud and
\teff\ $/$ \logg\ consistent with evolution model predictions (820K $/$ 4.0)
matched the available observed photometry at 3-$\sigma$.  While only
CH4$_{s/l}$, $J$, $H$, $K_s$, and L$^\prime$ photometry where available in
2008, this earlier (chemical equilibrium) model remains consistent 
($\sim$ 3-$\sigma$) with all of the revised/new broad-band photometry (see Fig.
\ref{fig14}).  However, the true level of agreement between this model and
reality is revealed by the OSIRIS data (Fig.  \ref{fig14}) where much weaker
CH$_4$ absorption is observed than predicted by this chemical equilibrium
model. This discrepancy highlights the difficulties in obtaining a
reliable description of young giant planets from photometry alone, which can
hide significant clues about their atmospheric properties.

To explore differences between the atmosphere and evolution derived bulk
parameter more closely, a separate grid of atmosphere models was computed, with
\teff\ and \logg\ uniformly sampling the evolution track predictions that are
consistent with the observed luminosity of \hrb\ (down the center of the grey
band in Fig.  \ref{fig13}). The clouds and non-equilibrium chemistry were
modeled in the same manner as in the solar abundance grid described above, yet
only \azero\ = 5 and 10 $\mu$m were used.  The metallicity, however, was
extended beyond solar abundances to include two metal rich cases ([Fe/H] = 0.5
and 1.0) and an abundance pattern matching the observed peculiar
($\lambda$-boo-type) subsolar abundances of HR8799 \cite[]{Sadakane2006}.  The
best fitting model from this ``evolution-consistent" grid has \teff\ = 896K,
\logg\ = 4.3, [Fe/H] = 1.0, $\log L_{\rm bol}/L_{\odot} = -5.06$ (see Fig.
\ref{fig14}).  The agreement between this metal-rich model and the OSIRIS $H$
and $K$-band spectra is similar to that of the previous hotter, solar
abundance, best fit.  The broad-band photometric comparison is also good (mostly to
1-$\sigma$), but slightly worse compared to the solar abundance model at $J$
and $K$ bands (see Figure \ref{fig14}) but still consistent at 2-$\sigma$.
The lowering of \teff\ by increasing metallicity is to be expected.  Looking at
the $M_J$ versus $J-K$ CMD in Fig. \ref{fig6}, in order for cooler
atmospheres to match \hrb\ they need to be redder in $J-K$ and this can be
accomplished by increasing metallicity and/or by decreasing the modal grain
size \cite[]{Burrows2006}.  The mean $\chi^2$ for this metal-rich comparison is
$\sim 5.5$ and only marginally better than the solar abundance model at the
same temperature and gravity. However, given that even in this smaller grid the
parameter space is still sparsely sampled, there is plenty of room for
improvement. 

\begin{figure}[!tb]
\plotone{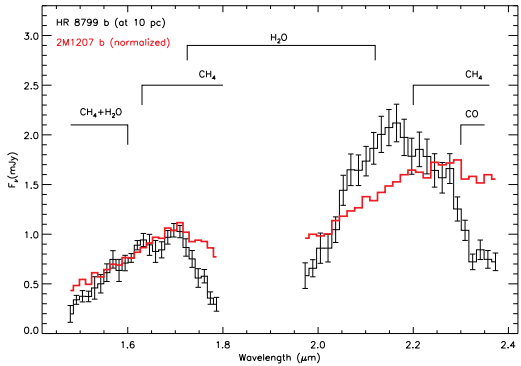}
\caption{
OSIRIS $H$ and $K$-band spectra of \hrb\ compared to the spectra of \twom\
\cite[]{Patience2010} binned to the same resolution and interpolated onto the
same wavelength grid as the OSIRIS observations.  The \twom\ spectra where
independently normalized for this comparison. The error-bars for the \twom\
spectra are significantly smaller than those of \hrb\ and, for clarify, are not
show.
\label{fig15}}
\end{figure}

\subsection{Comparison to 2M1207~b}

In near-IR CMDs, the object most similar to \hrb\ is \twom, a
planet-mass companion to a young ($\sim 10$ Myr) brown dwarf
\cite[]{Chauvin2005}.   In addition to their very red colors, these two objects
also share the same puzzling fact that their SEDs appear hotter than evolution
models predict, for their ages and luminosities.  Available photometry and
spectroscopy of \twom\ show all the signs of an object with effective
temperature $\sim 1600$K \cite[]{Mohanty2007, Ducourant2008, Patience2010}.
When combined with current luminosity estimates ($\log(L/L_\odot) \sim -4.5$),
the radius implied by this high effective temperature is $\sim 0.7$ \rjup,
comparable to the small size inferred above for \hrb.  Consequently, {\em both}
\twom\ and \hrb\ are underluminous compared to hot-start evolution models
and overluminous for cold start models, for their respective ages.

\cite{Mohanty2007} concluded that the high effective temperature was likely
correct and, thus, \twom\ is underluminous by $\sim 2.5$ magnitudes.  Given
that the distance to the source (52.4 $\pm$ 1.1 pc) is known
\cite[]{Ducourant2008}, \cite{Mohanty2007} proposed that an edge-on disk
surrounds \twom\ providing the necessary extinction.  Alternatively,
\cite{Mamajek2007} proposed that \twom\ is not a normal gas giant but rather
the product of a recent protoplanetary collision and the observed flux is the
afterglow from the hot, physically small, remnant.  Given that \hrb\
suffers from a similar quandary, it is unlikely that such rare events should be
necessary to explain the first and only two objects found in the very red and faint
regions of near-IR CMDs.

Given the puzzling similarities between \twom\ and \hrb\ mentioned above,
it is useful to compare their SEDs.  Moderate resolution ($R \sim 500 - 1500$)
$H$ and $K$-band spectra of \twom\ \cite[]{Patience2010} are compared to the
\hrb\ spectra in Figure \ref{fig15}.  The spectrum of \twom\ has
shallower H$_2$O bands with a distinct CO band-head at 2.3 $\mu$m and no
prominent CH$_4$ absorption -- a spectrum very much like typical L dwarfs.
The only feature the two sets of spectra have in common is a triangular shaped
spectrum in the $H$ band, consistent with both objects having low surface
gravity (indicative of low mass and/or youth).  Assuming both objects are not
the result of recent protoplanetary collisions, \twom\ and \hrb\ may well
represent important evolutionary states of substellar atmospheres. Independent
of the edge-on disk hypothesis, if the ages of the two systems are correct,
then the planets' masses are probably the same to within a factor of 2.  If the
masses are similar, then these objects provide evidence for rapid spectral
changes from 10 to $\sim 30$ Myr for young planet-mass objects.

The analysis of \hrb\ described above provides a third possible explanation
for the properties of \twom\ -- namely that certain combinations of cloud
coverage and non-equilibrium chemistry (and perhaps metallicity) are capable of
producing a very low temperature object with an L-type near-IR spectrum.  The
two hypotheses proposed by Mohanty et al. and Mamajek et al. hinge on the
assumption that the effective temperature of \twom\ is $\sim$ 1600K.
However, looking back at Figure \ref{fig6}, \twom\ is intersected by an
intermediate cloudy model with \teff\ = 1100K, close to evolution model
predictions and consistent with a typical radius.  Also, given the apparently
extreme reduction of CH$_4$ in the atmosphere of \hrb, it is conceivable
that the CO/CH$_4$ mole fractions are quenched at even higher pressures in the
younger (and likely lower gravity) atmosphere of \twom, resulting in a low
temperature atmosphere that is rich in CO (perhaps enough to produce a distinct
CO bandhead at 2.3 $\mu$m, as seen in the spectrum from Patience et
al.\nocite{Patience2010}) and without significant amounts of CH$_4$.  A search
for such a cold, cloudy, model atmosphere with extreme non-equilibrium
chemistry for \twom\ is underway and will be presented in a separate paper
(Barman et al., in prep.). 

Despite their spectral differences, the most straightforward explanation for
the positions of \twom\ and \hrb\ in near-IR CMDs is one of atmospheric
origins.  Instead of an edge-on disk or recent protoplanetary collision, a full
exploration of the cloud properties, non-equilibrium chemistry, and metallicity
is likely required to reproduce their observed SEDs with a model atmosphere
having \teff\ and \logg\ consistent with evolution predictions.  However, as
more and more important physical processes are identified in substellar
atmospheres, the number of free (and essentially independent) parameters
increases and the ability to exhaustively explore all possible parameter
combinations diminishes -- more work (along the lines of Freytag et al.
2010\nocite{Freytag2010}) is clearly required to reduce the current set of
parameters and produce more unified model atmospheres.

\section{Conclusions}

The HR8799 planetary system provides a rare opportunity to study multiple,
coeval, giant planets at very young ages.  The data and model comparisons
described above are only first steps toward understanding just one of these
planets and many challenging problems remain.  While the data presented here
offer the broadest spectral coverage for an HR8799 planet obtained so far, the
low resolution and modest signal-to-noise set strong limits on what can be
understood about \hrb.  In addition to the limits imposed by the data, the
complexity and excess of free parameters in model atmospheres for giant planets
and brown dwarfs also set strong limits on what can be confidently inferred
from observations. With these limitations in mind, several conclusions
can be drawn from the observations of \hrb.

The SED of \hrb\ is shaped by water absorption bands that are deeper than
typically seen in late L and early T field dwarfs, but clearly consistent with
a Hydrogen-rich planetary atmosphere.  Only weak methane absorption is detected
in the $K$ band along with very weak CO absorption, indicating strong
non-equilibrium chemistry and vertical mixing.  Additional evidence for very
weak methane absorption is seen in the $H$ band spectrum (and narrow-band
photometry).  The near-IR spectra and broad-band photometry are smoothed and
reddened by photospheric clouds not typical for brown dwarfs with \teff $\le$
1100K.   The triangular shape of the $H$-band spectrum clearly indicates low
surface gravity ($\log(g) < 5$), consistent with an age significantly younger
than that proposed by \cite{Moya2010}.  

The best match from a solar abundance grid of model atmospheres (having \teff =
1100K and \logg = 3.5) found above implies a radius that is too small to be
consistent with both the observed luminosity and the current understanding of
the interiors and evolution of H+He-rich planets, regardless of mass, age, or
interior core-size.  Furthermore, a variety of model atmospheres can be
constructed that fit the observed broad-band photometry very well (at least
from the model grids used here), yet do not match the observed spectra --
indicating that parameters inferred from broad-band photometric fits alone
should be treated with caution.

While certainly not perfect, the model comparisons presented here are better
representations of the observations than those presented by \cite{Bowler2010}
and \cite{Currie2011}.  In particular, Bowler et al. found fits to their
narrow-band spectrum that require \teff $> 1100$K,  which leads to unphysically
small radii of $R < 0.75$\rjup. These previous studies also had difficulty
simultaneously fitting the photometry; for example, Currie et al.  found it
difficult to match most of the photometry, especially at $\lambda > 3\mu$m
(even with their thick or patchy cloud models) and Bowler et al. had difficulty
matching the near-IR colors for reasonable \teff.  These difficulties may stem
from the sparse sampling of model parameters in the grids used in these two
studies (e.g. those parameters regulating non-eq. chemistry and clouds), though
considerable degeneracies are present when fitting only broad-band photometry.
Furthermore, the best fitting models for \hrb\ found by Currie et al.
typically have deeper methane absorption than seen in the broad $K$-band
spectrum described above or the narrow-band spectrum of Bowler et al.
Interestingly, Currie et al.  emphasize cloud thickness/coverage over
non-equilibrium chemistry, in particular as it pertains to the near-IR SED.
The analysis presented here, however, requires deep quenching of CO/CH$_4$ to
match the near-IR spectra (and near-IR narrow-band photometry), placing equal
importance on both atmospheric effects when reproducing the observations.
 
Finding a solar abundance atmosphere that matches the complete set of data and
interior/cooling-track predictions seems unlikely (though still not
inconceivable). A more likely scenario is that the atmosphere of \hrb\ has an
elemental composition different from the Sun.  A model atmosphere that matches
the observations reasonably well (as well or better than any previous study),
including the predictions of evolution models, can be found if the atmosphere
is enhanced by a factor 10 in metals compared to the Sun.  Such a model has
\teff\ = 896K and \logg\ = 4.3, consistent with a young giant planet with mass
less than 10 \mjup.  At $\sim$ 30 Myrs, as recently proposed by
\cite{Zuckerman2011}, the mass of \hrb\ would be 4 to 5 \mjup.  
 
If \hrb\ is indeed metal rich and near the very low end of the original mass
range inferred by \cite{Marois2008}, then core-accretion may be the preferred
formation mechanism for the HR8799 planetary system.  Metal enrichment,
relative to the host star, is often considered a hallmark of the core-accretion
scenario and recent work indicates that if the HR8799 planets formed by
gravitation instabilities,  they should show little metal enhancement
\cite[]{Helled2010}.  Atmosphere models with either solar abundances or the
metal poor $\lambda$-boo-like abundances of the star result in \teff\ that are
too high and corresponding radii that are too small.  The analysis of
\cite{Bowler2010} also indicates that enhanced metals (and low gravity) are
likely to produce the best model fits.

While these conclusions rest heavily on models not tested by a large population
of young giant planets (such a population should emerge in the coming years),
the conclusions are self-supporting.  Low surface gravity, clearly preferred by
the model fitting (regardless of metallicity) and indicated by the shape of the
$H$-band spectrum, may well allow for more efficient vertical mixing.  As
surface gravity decreases, the radiative-convection boundary moves closer to
the photosphere, the maximum convective velocity increases, and \kzz\ also
increases, at least in the convection zone \cite[]{Freytag2010,Rice2011}.
Enhanced vertical mixing leads to enhanced non-equilibrium chemistry, as
suggested by the absence of CH$_4$ absorption in both $H$ and $K$ bands.
However, the connection between gravity and mixing is merely suggestive as the
primary physical mechanism for mixing into the radiative zone unknown.  Low
surface gravity may also establish atmospheric conditions (not found in the
higher gravity field T dwarfs) that support thick clouds at photospheric
depths, allowing the intersection of the T-P curve and the condensation curve
to be near the photosphere even at low \teff. A connection between clouds and
surface gravity is already starting to emerge for brown dwarfs.  Evidence for a
gravity dependence for the L-to-T transition has already been reported
\cite[]{Metchev2006, Saumon2008} and, recently, \cite{Stephens2009} found that
low-gravity brown dwarfs in their sample remain cloudy at lower effective
temperatures, with the L-to-T transition temperature decreasing from 1300K to
1100K as \logg\ decreases from 5.0 to 4.5.  \cite{Metchev2006} also found that
HD20303B, a low gravity transition brown dwarf, is $\sim$ 250K cooler than
field objects with similar spectral types.  If \hrb\ has a \logg\ $\sim 4$,
then it is possible that this emerging trend extends to this object as well,
and would suggest that very low gravity objects (like young giant planets) are
cloudy down to temperatures well below 1000K.  The structure of the clouds
needed to reproduce the colors and spectra of the HR8799 planets and 2m1207b
are similar to those needed for late L-dwarfs. This suggests that such objects,
from an atmospheric point of view, are extensions of the L-type field dwarfs
rather than members of a new class altogether.  This conclusion is distinct
from that of Currie et al. who claim the clouds must be significantly thicker
than those required to match L dwarfs.  Finally, enhanced metals would also
lead to enhanced grain formation and conditions ripe for atmospheric cloud
formation.  The combination of these atmospheric conditions lead to bulk
properties, inferred from the SED analysis alone, that include a radius
consistent with the observed luminosity and current understanding of giant
planet interiors, as well as a mass that fits comfortably within the limits
required for long-term dynamical stability.

\acknowledgements
We thank the referee, Jonathan Fortney, for his careful review of our paper.
Additional thanks go out to Shelley Wright, Jim Lyke, and James Larkin for
their assistance with all things OSIRIS-related.  We also thank Brad Hansen,
Mark Marley, Didier Saumon, and Tristan Guillot for many useful discussions on
the vagaries of planets and brown dwarfs.  We are indebted to Jim Lyke, Al
Conrad, Hien Tran, Scott Dahm, Randy Campbell, Terry Stickel and the entire
Keck staff for their assistance in maximizing our observing efficiency.  TB
acknowledges the Kavli Institute for Theoretical Physics and the participants
of the 2010 workshop on Exoplanets for providing a stimulating environment.
This research has benefited from the SpeX Prism Spectral Libraries, maintained
by Adam Burgasser.  The data presented herein were obtained at the W.M. Keck
Observatory, which is operated as a scientific partnership among the California
Institute of Technology, the University of California and the National
Aeronautics and Space Administration.  The Observatory was made possible by the
generous financial support of the W.M.  Keck Foundation. The authors wish to
recognize and acknowledge the very significant cultural role and reverence that
the summit of Mauna Kea has always had within the indigenous Hawaiian
community.  We are most fortunate to have the opportunity to conduct
observations from this mountain.  Most of the numerical work was carried out at
the NASA Advanced Supercomputing facilities.  This research was support by 
NASA Origins of Solar Systems grants to LLNL and Lowell Observatory.  This
research was also support by JPL/NexSci RSA awards.  We thank all these
institutions for their support.

\end{document}